\documentclass[superscriptaddress,secnumarabic,twocolumn,
 amssymb,amsmath,nobibnotes,aps,prd,showkeys,showpacs]{revtex4}
\usepackage{graphicx,epsfig,amssymb,amsmath,amstext}
\usepackage{bm}
\usepackage{booktabs}
\usepackage{color}
 
\newcommand{\pd}[2]{\frac{\partial #1}{\partial #2}} 
 
\newcommand{\grad}[1]{\mathbf{\nabla} #1} 
\let\baraccent=\= 
\renewcommand{\=}[1]{\stackrel{#1}{=}} 

\begin{document}
\title{Investigating spinning test particles: spin supplementary conditions and
the Hamiltonian formalism }

\author{Georgios Lukes-Gerakopoulos}
\email{gglukes@gmail.com}
\affiliation{Theoretical Physics Institute, University of Jena, 07743 Jena,
Germany}
\author{Jonathan Seyrich}
\email{seyrich@na.uni-tuebingen.de}
\affiliation{Mathematisches Institut, Universit\"{a}t T\"{u}bingen,
Auf der Morgenstelle, 72076 T\"{u}bingen, Germany}
\author{Daniela Kunst}
 \email{daniela.kunst@zarm.uni-bremen.de}
\affiliation{ZARM, University of Bremen, Am Fallturm, 28359 Bremen, Germany}

\begin{abstract}
 In this paper we report the results of a thorough numerical study
 of the motion of spinning particles in Kerr spacetime with different prescriptions.
 We first evaluate the Mathisson-Papapetrou equations with two different spin
 supplementary conditions, namely, the Tulczyjew and the Newton-Wigner, and make a
 comparison of these two cases. We then use the Hamiltonian 
 formalism given by Barausse, Racine, and Buonanno in 
 [{\it Phys. Rev. D} {\bf 80}, 104025 (2009)] to evolve the orbits and compare them
 with the corresponding orbits provided by the Mathisson-Papapetrou equations. We
 include a full description of how to treat the issues arising in the numerical
 implementation.
\end{abstract}
%
%

\maketitle

\section{Introduction}\label{sec:Intro}

 Since we expect that the centers of galaxies are occupied by supermassive black
 holes, relativistic binary systems with extreme mass ratios are of great interest.
 A first approximation to an extreme mass ratio inspiral (EMRI) is the geodesic 
 motion where the spin of the smaller particle is ignored.  More relevant
 models have to incorporate the spin. This, however, appears not to be so simple.

 The equations of motion of a spinning particle were given by 
 Mathisson \cite{Mathisson37} and Papapetrou \cite{Papapetrou51} several decades
 ago. The Mathisson-Papapetrou (MP) equations are not a closed set of first order 
 ordinary differential equations, i.e., there are less equations than necessary in
 order to  evolve the system. To close the set, an extra spin supplementary
 condition (SSC) is required. Over the years, various such SSCs have been proposed
 (see, e.g., \cite{Semerak99,Kyrian07} for a review).

 As a SSC fixes a center of reference, e.g., the center of the mass, and different
 SSCs define different centers, for each SSC we have a different world line (see,
 e.g., \cite{Kyrian07}), and, hence, each SSC prescribes a different evolution of
 the MP equations. But, although this ambiguity appears to be a major issue in the
 modeling of an EMRI binary system, the difference in the evolution caused by
 different SSCs has  not received the adequate attention. Our work aims at
 quantifying those evolution differences in a Kerr spacetime background.

 The first part of the study addresses the above issue by examining how ``similar''
 initial conditions diverge when they are evolved by using different SSCs. We focus
 on two SSCs, namely the Tulczyjew (T) SSC  \cite{Tulczyjew59} and the
 Newton-Wigner(NW) SSC \cite{NewtonWigner49}, as introduced by Barausse et al. in
 \cite{Barausse09}. T~SSC is a standard SSC that has been used in several works
 concerned with different topics, see, e.g.,
\cite{Semerak99,Kyrian07,Dixon70,Mino96,Hartl03,Steinhoff12,Bini14,Ramirez14,Hackmann14}.
 On the other hand, NW~SSC has been successfully implemented in the framework of the
 Post-Newtonian approximation \cite{Steinhoff08,Steinhoff11}, and it is the only
 SSC allowing for a canonical Hamiltonian formalism, albeit only up to linear
 order in the spin of the particle in curved spacetimes. This Hamiltonian formalism
 has been  derived in~\cite{Barausse09}. As it has many practical advantages to
 have a Hamiltonian formulation of a given problem at hand, for example because it
 is part of the effective one body formulation \cite{Barausse09,Barausse10}, it
 would be nice to see if orbits obtained via the Hamiltonian formalism
 of~\cite{Barausse09} stay close to those obtained with the help of the full MP 
 equations in the case of NW SSC (a discussion on the topic can be found in Sec. IV
 of \cite{Hinderer13}). Therefore, in the second part of our work, we compare both
 approaches numerically. 
 
 A numerical investigation of the equations considered in this work entails a bunch 
 of interesting numerical challenges. To start with, a useful study of the
 divergence of different orbits should straddle a reasonably long time interval.
 The efficient integration of equations of motion over a long time interval
 requires structure preserving algorithms (see, e.g.,~\cite{hairerlubichwanner} for
 an elaborate overview) such as symplectic schemes, which have been successfully
 applied for simulations in various fields of general relativity,
 e.g.,~\cite{Seyrich12,Seyrich13,zhongwu10,wuetal13}. Moreover, the MP equations
 have no Hamiltonian structure, wherefore one would expect usual symplectic integration 
 schemes to lose their theoretical advantage over ordinary, not so efficient ones. 
 What is more, in the NW~SSC case part of the equations of motion will turn out to be
 known only implicitly. In this work we will explain how, notwithstanding the just
 mentioned obstacles, the MP equations can be evolved accurately in an efficient way
 for both SSCs. When comparing orbits calculated via the MP equations with those
 obtained by the Hamiltonian equations of~\cite{Barausse09}, one is faced with the
 problem of different evolution parameters. We thus come up with a comfortable way
 of guaranteeing output at consistent times.

 The paper is organized as follows. In Sec.~\ref{sec:MPeqs} we introduce the MP
 equations and give a brief discussion on the SSCs. Then, we turn to the Hamiltonian
 formalism in  Sec.~\ref{sec:HamSP}, where the basic elements concerning the
 Hamiltonian function, which describes the  motion of a spinning particle in curved
 spacetime, are summarized. In Sec.~\ref{sec:TNWCom} we explain how the simulations
 with the MP equations are done, and a comparison between the T and the NW~SSC is 
 provided, whilst Sec.~\ref{sec:MPHamCom} quantifies the difference in the
 evolution of orbits between the MP equations and their Hamiltonian approximation.
 Finally, we discuss our main results in  Sec.~\ref{sec:con}. A detailed
 discussion of the numerical implementation is provided in the Appendix 
 (Secs.~\ref{sec:NumIntMP}~and~\ref{sec:NumIntHam}).

 The units we use are geometric $(G=c=1)$, and the signature of the metric is
 (-,+,+,+). Greek letters denote the indices corresponding to spacetime (running
 from 0 to 3), while Latin ones denote indices corresponding only to space (running
 from 1 to 3). We use capital letters for the indices when referring to a flat
 spacetime. In general, we try to follow the notation in \cite{Barausse09} whenever
 this is possible.

\section{Mathisson-Papapetrou equations}\label{sec:MPeqs} 

 The Mathisson-Papapetrou equations describe the motion of a particle with mass
 $\mu$ and spin $S^{\mu\nu}$ (pole-dipole approximation) in a given background
 $g_{\mu\nu}$. Their formulation in \cite{Dixon70} reads
 \begin{eqnarray}
  \frac{D~p^\mu}{d \tau}=-\frac{1}{2}~{R^\mu}_{\nu\kappa\lambda}
  v^\nu S^{\kappa\lambda}~~,
  \label{eq:MPmomenta} \\
  \nonumber \\
  \frac{D~S^{\mu\nu}}{d \tau}=p^\mu~v^\nu-v^\mu~p^\nu~~, \label{eq:MPspin}
 \end{eqnarray}
 where $p^\mu$ is the four-momentum, $v^\mu=d x^\mu/d \tau$ is the tangent
 vector to the worldline along which the particle moves, $\tau$
 is the proper time along this worldline, and ${R^\mu}_{\nu\kappa\lambda}$ is the
 Riemann tensor. In the case of a stationary and axisymmetric  spacetime, the energy
  \begin{align}
   E &= -p_t+\frac12g_{t\mu,\nu}S^{\mu\nu}~~,\label{eqn-constant-of-E}
  \end{align}
  and the $z$ angular momentum 
  \begin{align}
  J_z &= p_\phi-\frac12g_{\phi\mu,\nu}S^{\mu\nu}~~,\label{eqn-constant-of-J_z}
  \end{align}
  are preserved along the solutions of the MP equations.

 Since we selected $\tau$ to be the proper time, it holds that $v^\nu~v_\nu=-1$.
 By multiplying Eq.~(\ref{eq:MPspin}) with $v_\nu$ we get  
 \begin{equation}
  p^\mu=m~v^\mu-v_\nu~\frac{D~S^{\mu\nu}}{d\tau}~~, \label{eq:MomTang}
 \end{equation}
 where $m=-p^\nu~v_\nu$ is the rest mass of the particle with respect to
 $v^\nu$, while the measure of the four-momentum $p^\nu~p_\nu=-\mu^2$ provides the 
 rest mass $\mu$ with respect to $p^\mu$. $m=\mu$ holds only if the tangent vector
 $v^\nu$ coincides with the four-velocity $u^\nu=p^\nu/\mu$.

 It is useful to stress that neither of the masses have to be a constant of motion.
 Namely, for $m$ we get
 \begin{equation}
  \frac{d m}{d \tau}=\frac{D~m}{d \tau}=-\frac{D~v_\nu}{d \tau}~p^\nu~~, \nonumber
 \end{equation}
 since from Eq. (\ref{eq:MPmomenta}) we see that
 $\displaystyle\frac{D~p^\nu}{d \tau} v_\nu=0$, and by using Eq.~(\ref{eq:MomTang})
 for replacing $p^\nu$, we arrive at
 \begin{equation}
  \frac{d m}{d \tau}=\frac{D~v_\nu}{d\tau}~v_\mu~\frac{D~S^{\nu\mu}}{d\tau}~~. 
  \label{eq:MomMassCon}
 \end{equation}
 For $\mu$ we have
 \begin{equation}
  \frac{d \mu}{d \tau}=\frac{D~\mu}{d \tau}=-\frac{p_\nu}{\mu}~\frac{D~p^\nu}{d \tau}~~, \nonumber
 \end{equation}
 and again by using Eq.~(\ref{eq:MomTang}) for replacing $p^\nu$, we get
 \begin{equation}
  \frac{d \mu}{d \tau}=\frac{D~p_\nu}{d\tau}~\frac{p_\mu}{\mu~m}
 ~\frac{D~S^{\nu\mu}}{d\tau}~~. 
  \label{eq:EffMassCon}
 \end{equation}

 The same holds for the spin measure 
 \begin{equation}
   S^2=\frac{1}{2}~S_{\mu\nu}~S^{\mu\nu}~~. \label{eq:SpinMeasure}
 \end{equation}
 Here, we have
 \begin{equation}
   \frac{d~S^2}{d\tau}=\frac{D~S^2}{d\tau}=S_{\mu\nu}~\frac{D~S^{\mu\nu}}{d\tau}~~, \label{eq:SpinCoA}
 \end{equation} 
 and by Eq.~(\ref{eq:MPspin}) we get
 \begin{eqnarray}
   \frac{d~S^2}{d\tau} &=& S_{\mu\nu}~(p^\mu~v^\nu-v^\mu~p^\nu) \nonumber \\
                       &=& 2 S_{\mu\nu}~p^\mu~v^\nu~~, \label{eq:SpinCoB}
 \end{eqnarray} 
 which becomes zero if
 \begin{equation}
   S_{\mu\nu}~p^\mu=0~~, \label{eq:Tulczyjew}
 \end{equation}  
 or
 \begin{equation}
   S_{\mu\nu}~v^\mu=0~~. \label{eq:Pirani}
 \end{equation}  

 Eq.~(\ref{eq:Tulczyjew}) is the Tulczyjew SSC, while Eq.~(\ref{eq:Tulczyjew}) is
 the Pirani SSC \cite{Pirani56}. From Eq.~(\ref{eq:EffMassCon}) we see that
 $d\mu/d\tau=0$ for T~SSC, while for Pirani SSC $dm/d\tau=0$. The MP equations with
 Pirani SSC exhibit a ``strange'' helical motion (see, e.g., \cite{Kyrian07}), which
 has been considered as unphysical. However, recently, in \cite{Costa12} the authors
 argued that the  helical motion can be interpreted by the concept of a hidden
 electromagnetic-like momentum. We will not discuss Pirani SSC further. Instead, we
 are going to focus on the Newton-Wigner SSC, which reads
 \begin{equation}
   S^{\mu\nu}~\omega_\mu=0~~, \label{eq:NW}
 \end{equation} 
 where $\omega_\mu$ is a time-like vector, or a sum of time-like vectors, e.g.,
 of $p_\mu$ and $\varphi_\mu$, i.e.,
 \begin{equation}
 \omega_\mu=p_\mu+\mu~\varphi_\mu~~. \label{eq:NWvect}
 \end{equation} 

 In general, for NW~SSC, neither the masses, Eqs.~(\ref{eq:MomMassCon}), 
 (\ref{eq:EffMassCon}), nor the spin, Eq.~(\ref{eq:MPspin}), are preserved. Thus,
 from this point of view it is a strange selection of a SSC. However, we should
 keep in mind that our framework is a pole-dipole approximation. Therefore it is
 somehow adequate for the just mentioned quantities to be conserved only up to
 linear order in the spin. For the spin, this can be seen from
 Eq.~(\ref{eq:SpinCoA}) but for the mass $\mu$ the proof is quite more complicated
 and was provided in \cite{Barausse09}.

 \subsection{Spin four-vector} \label{subsec:S4V}
 
 Instead of the spin tensor $S^{\mu\nu}$, a spin four-vector $S^\mu$ is used 
 sometimes, since $S^\mu$ is often considered more physically
 intuitive and more convenient than $S^{\mu\nu}$ (see, e.g., \cite{Suzuki97}).

 For the T~SSC the four-vector is defined by
 \begin{equation}
   S_{\mu}=-\frac{1}{2} \eta_{\mu\nu\rho\sigma} u^\nu S^{\rho\sigma}~~, \label{eq:T4VS}
 \end{equation}
 where $\eta_{\mu\nu\rho\sigma}$ is the Levi-Civita density tensor 
 \begin{equation} \label{eq:LevCivD}
   \eta_{\mu\nu\rho\sigma}=\sqrt{-g}~\epsilon_{\mu\nu\rho\sigma}~~,
 \end{equation}
 and $\epsilon_{\mu\nu\rho\sigma}$ is the Levi-Civita symbol with
 $\epsilon_{0123}=-1$. The inverse relation of Eq.~(\ref{eq:T4VS}) between the two
 spin forms is
 \begin{equation}
   S^{\rho\sigma}=-\eta^{\rho\sigma\gamma\delta} S_{\gamma} u_\delta~~. \label{eq:T4VSin}
 \end{equation}
 By replacing the last equation in Eq.~(\ref{eq:SpinMeasure}), we get
 \begin{equation}
   S^2=S_{\mu}~S^{\mu}~~. \label{eq:TSpinMeas4V}
 \end{equation}
 From Eq.~(\ref{eq:T4VS}) we see that
 \begin{equation}
   S_{\mu} p^\mu=0~~, \label{eq:T4VOr}
 \end{equation}
 so the spin four vector is perpendicular to the momentum.
 
 For the NW~SSC we define the four-vector as
 \begin{equation}
   S_{\mu}=-\frac{1}{2~\mu} \eta_{\mu\nu\rho\sigma} \omega^\nu S^{\rho\sigma}~~.
  \label{eq:NW4VS}
 \end{equation}
 By this definition we fix that 
 \begin{equation}
   S_{\mu} \omega^\mu=0~~. \label{eq:NW4VOr}
 \end{equation}
 Thus, the spin four vector is perpendicular to the time-like vector $\omega_\mu$.
 In the NW case the inverse relation of Eq.~(\ref{eq:NW4VS}) between the two spin
 forms is 
 \begin{equation}
   S^{\rho\sigma}=\eta^{\rho\sigma\gamma\delta}~S_{\gamma}~
 \frac{\mu~\omega_\delta}{\omega_\nu \omega^\nu}~~. \label{eq:NW4VSin}
 \end{equation}
 Now, the spin measure (\ref{eq:SpinMeasure}) reads
 \begin{equation}
   S^2=-\frac{\mu^2}{\omega_\nu \omega^\nu}~S_{\sigma}~S^{\sigma}~~. \label{eq:NWSpinMeas4V}
 \end{equation}

 The measure of the spin divided by the rest mass, i.e., $S/\mu$ defines the 
 minimal radius of a volume which a spinning body has to have in order not to
 rotate with superluminal speed. The same radius defines the upper bound of the 
 separation between worldlines defined by various SSCs, i.e., a disc of centers of
 mass inside of which the worldlines have to lie. This radius was introduced by
 M\"{o}ller in \cite{Moller49} and, therefore, is often referred to as the
 M\"{o}ller radius. 

 In the next step, we explain how to calculate the tangent vector $v^\mu$.

 \subsection{Calculating the tangent vector}  \label{subsec:TanV}

 The MP equations do not explicitly state how we can evaluate the tangent vector
 $v^\mu$ throughout the evolution. To find $v^\mu$ we use the SSCs.

 In the case of T~SSC, $v^\mu$ is found via the relation
 \begin{equation}
  v^\mu=N (u^\mu+w^\mu)~~,\label{eq:TTanVA}
 \end{equation}
 where
 \begin{equation}
  w^\mu=\frac{2~S^{\mu\nu}~u^\lambda R_{\nu\lambda\rho\sigma}~S^{\rho\sigma}}
  {4~\mu^2+R_{\alpha\beta\gamma\delta}~S^{\alpha\beta}~S^{\gamma\delta}}~~,
 \label{eq:TTanVB}
 \end{equation}
 and, because $v^\mu v_\mu=-1$, we get
 \begin{equation}
  N=\frac{1}{\sqrt{1-w_\mu~w^\mu}}~~.\label{eq:TTanVC}
 \end{equation}
 For more details on how to derive the above expression see, e.g., \cite{Semerak99}.

 In the case of NW~SSC, according to our knowledge, there is no explicit expression
 which gives $v^\mu$ as a function of $p^\mu$ and $S^{\mu\nu}$. However, by
 taking the covariant derivative of Eq.~(\ref{eq:NW}), we obtain
 \begin{equation}
  v^\mu=\frac{1}{\omega_\nu p^\nu}\left( (\omega_\nu v^\nu) p^\mu+S^{\mu\nu}
 \frac{D~\omega_\nu}{d\tau}  \right)~~. \label{eq:NWTanV}
 \end{equation}
 A detailed discussion on how we solve the initial value problem numerically
 is provided in Appendix~\ref{sec:NumIntMP}.

\section{The Hamiltonian formalism for the spinning particle}\label{sec:HamSP}

 The MP equations (\ref{eq:MPmomenta}),~(\ref{eq:MPspin}) can be derived by means 
 of Lagrangian mechanics, see, e.g.,~\cite{Westpfahl69,Bailey75,Porto06}. 
 If we want to apply a Legendre transformation in order to get a Hamiltonian
 canonical formulation \footnote{There is also another Hamiltonian formulation
 for the spinning particle \cite{Ramirez14} in which a noncommutative position
 coordinate is used instead of the canonical one.} for a spinning particle moving
 in a curved spacetime, then the canonical structure holds only at linear order
 of the particle's spin \cite{Barausse09}. 

 The spin in the Hamiltonian formalism proposed by \cite{Barausse09} comes from the
 projection of the spin tensor $S^{\mu\nu}$ onto the spacelike part of a tetrad
 field $\tilde{e}^\mu_\Delta$. This tetrad consists of a timelike future oriented
 vector $\tilde{e}^\mu_T$ (throughout the article we shall use T instead of 0) and
 three spacelike vectors $\tilde{e}^\mu_I$. For the tetrad it holds that
 \begin{equation}
  \tilde{e}^\mu_{~\Gamma} \tilde{e}^\nu_{~\Delta}~g_{\mu\nu}= \eta_{\Gamma\Delta}~~,
 \label{eq:tetradflat}
 \end{equation}
 where $\eta_{\Gamma\Delta}$ is the metric of the flat spacetime, and
 \begin{equation}
  \tilde{e}^\mu_{~\Delta} \tilde{e}_\nu^{~\Delta}=\delta^\mu_\nu~~,
 \label{eq:tetradcom}
 \end{equation}
 where $\delta^\mu_\nu$ is the Kronecker delta. The capital indices are raised or
 lowered by the flat metric.  When a tensor is denoted with capital indices, then 
 the tensor has been projected onto this tetrad $\tilde{e}^\mu_\Delta$. In the case
 of the spin tensor $S^{\mu\nu}$, the projection reads
 \begin{equation}
  S^{IJ}=S^{\mu\nu}~\tilde{e}_\mu^{~I}~\tilde{e}_\nu^{~J}~~. \label{eq:SpinProj}
 \end{equation}
 The remaining components of this projection come from splitting the
 NW~SSC~(\ref{eq:NW}) appropriately, and projecting the split on the tetrad, i.e.,
 \begin{equation}
  S^{TI}=S^{IJ}~\frac{\omega_J}{\omega_T}~~, \label{eq:SpinProjT}
 \end{equation}
 where $\omega_\Delta=\tilde{e}^\nu_{~\Delta}\omega_\nu$ is the projection of the 
 time-like vector (\ref{eq:NWvect}) of the NW~SSC (\ref{eq:NW}) as chosen 
 in~\cite{Barausse09}
 \begin{equation}
  \omega_\nu=p_\nu-\mu~\tilde{e}_\nu^{~T}~~ \label{eq:NWTimeLV}
 \end{equation}
 on the tetrad field, i.e.,
 \begin{eqnarray}
  \omega_T &=& p_\nu~\tilde{e}^\nu_{T}-\mu~~, \nonumber \\
  \omega_J &=& p_\nu~\tilde{e}^\nu_{J}~~. \label{eq:omega}
 \end{eqnarray}

 However, the Hamiltonian function of the spinning particle given
 in~\cite{Barausse09} does not use exactly the above described spin projection, 
 instead the spin three vector is employed, i.e.,
 \begin{equation}
  S_I=\frac{1}{2} \epsilon_{IJL}~S^{JL} \label{eq:3ProjSpin}
 \end{equation}
 (the inversion of Eq.~(\ref{eq:3ProjSpin}) gives $S^{JL}=-\epsilon^{JLI} S_I$).

 The Hamiltonian function $H$ itself 
 \begin{equation}
  H=H_{NS}+H^C~S_C~~, \label{eq:HamSP}
 \end{equation}
 splits in two parts. The first 
 \begin{equation}
  H_{NS}=\beta^{i}P_i+\alpha~\sqrt{\mu^2+\gamma^{ij}P_i P_j} \label{eq:HamNSP}
 \end{equation}
 is the Hamiltonian for a non-spinning particle, and the second $H^C~S_C$
 \begin{equation}
   H^C=-\left(\beta^{i} F_i^C+F_0^C+\frac{\alpha~\gamma^{ij}P_i~F_j^C}
 {\sqrt{\mu^2+\gamma^{ij}P_i P_j}}\right) \label{eq:HamSPc}
  \end{equation}
 includes the elements describing the spin, where 
 \begin{eqnarray}
  \alpha &=& \frac{1}{\sqrt{-g^{00}}}~~,\label{eq:alpha} \\
  \beta^i &=& \frac{g^{0i}}{g^{00}}~~, \label{eq:beta} \\
  \gamma^{ij} &=& g^{ij}- \frac{g^{0i}g^{0j}}{g^{00}}~~. \label{eq:gamma}
 \end{eqnarray}

 The canonical momenta $P_i$ conjugate to $x^{i}$ of the
 Hamiltonian~\eqref{eq:HamSP} can be calculated from the momenta $p_i$ of the MP
 formulation by using the relation
 \begin{eqnarray}
  P_i &=& p_i+E_{i\Gamma\Delta}S^{\Gamma\Delta}~~,\nonumber\\
  &=& p_i+\left(2 E_{iTJ}\frac{\omega_C}{\omega_T}+E_{iJC}\right)
 \epsilon^{JCL}~S_L~~,
  \label{eq:momentaHL}
 \end{eqnarray}
 where 
 \begin{equation}
   E_{\nu\Gamma\Delta}=-\frac{1}{2}\left(g_{\kappa\lambda}~
   \tilde{e}^\kappa_{~\Gamma}
   ~\frac{\partial\tilde{e}^\lambda_{~\Delta}}{\partial x^\nu}
  + \tilde{e}^\kappa_{~\Gamma}~\Gamma_{\kappa\nu\lambda}~
    \tilde{e}^\lambda_{~\Delta}\right)
  \label{eq:EmuAB}
 \end{equation}
 is a tensor which is antisymmetric in the last two indices, i.e.,
 $E_{\nu\Gamma\Delta}=-E_{\nu\Delta\Gamma}$. $\Gamma_{\kappa\nu\lambda}$, in turn,
 are the Christoffel symbols. This choice of momenta leads to a set of phase space
 variables that are canonical at linear order in the particle's spin.

 Finally, the $F_\mu^C$ tensor in Eq.~(\ref{eq:HamSP}) reads
 \begin{equation}
  F_\mu^C=\left(2 E_{\mu TI}\frac{\bar{\omega}_J}{\bar{\omega}_T}+E_{\mu IJ}\right)
 \epsilon^{IJC}~~, \label{eq:FmuC}
 \end{equation}
 where
 \begin{eqnarray}
  \bar{\omega}_\Delta &=& \bar{\omega}_\nu~\tilde{e}^\nu_{~\Delta}~~,\nonumber \\
  \bar{\omega}_\nu &=& \bar{P}_\nu-\mu~\tilde{e}_\nu^{~T}~~,\nonumber \\
  \bar{P}_i &=& P_i~~,\nonumber\\
  \bar{P}_0 &=& -\beta^i~P_i-\alpha\sqrt{\mu^2+\gamma^{ij}P_i P_j}~~, \nonumber\\
  \bar{\omega}_T &=& \bar{P}_\nu~\tilde{e}^\nu_{~T}-\mu~~, \nonumber \\
  \bar{\omega}_J &=& \bar{P}_\nu~\tilde{e}^\nu_{~J}~~.\label{eq:omegabar}
 \end{eqnarray}

 The equations of motion for the canonical variables as a function of coordinate
 time $t$, as derived in~\cite{Barausse09}, read
 \begin{align}
  \frac{d x^i}{dt} &=\pd H{P_i}~~,\label{eqn-Ham-eq-of-motion-x} \\
  \frac{d P_i}{dt} &=-\pd H{x^i}~~,\label{eqn-Ham-eq-of-motion-P}\\
 \frac{d S_I}{dt} &=\epsilon_{IJC}\pd H{S_J}S^C~~\label{eqn-Ham-eq-of-motion-S}~~.
 \end{align}

 The formulation provided up to this point is general, namely it does not depend on
 the coordinate or on the tetrad field choice. In the next section we specify the 
 setup we use in the numerical sections of our work.

 \subsection{The Hamiltonian for the Kerr spacetime} \label{subsec:HamKerrSP}

 The line element of the Kerr spacetime in Boyer-Lindquist coordinates is
 \begin{eqnarray}
  ds^2 &=& g_{tt}~dt^2+2~g_{t\phi}~dt~d\phi + g_{\phi\phi}~d\phi^2 \nonumber \\
       &+& g_{rr}~dr^2+g_{\theta\theta}~d\theta^2~~, \label{eq:LinEl}
 \end{eqnarray}
 where
 \begin{eqnarray}
   g_{tt} &=&-1+\frac{2 M r}{\Sigma}~~,\nonumber\\ 
   g_{t\phi} &=& -\frac{2 a M r \sin^2{\theta}}{\Sigma}~~,\nonumber\\
   g_{\phi\phi} &=& \frac{\Lambda \sin^2{\theta}}{\Sigma}~~, \label{eq:KerrMetric}\\
   g_{rr} &=& \frac{\Sigma}{\Delta}~~,\nonumber\\
   g_{\theta\theta} &=& \Sigma~~,\nonumber
 \end{eqnarray} 
 and
 \begin{eqnarray}
  \Sigma &=& r^2+ a^2 \cos^2{\theta}~~,\nonumber\\
  \Delta &=& \varpi^2-2 M r~~,\nonumber \\ 
  \varpi^2 &=& r^2+a^2~~, \nonumber \\ 
  \Lambda &=& \varpi^4-a^2\Delta \sin^2\theta~~.  \label{eq:Kerrfunc} 
 \end{eqnarray}
 In this section we reproduce the quantities already presented in \cite{Barausse09}.
 In the case of the small indices, we replace the numbers with the coordinates
 , i.e., $t,~r,~\theta,~\phi$ stand for $0,~1,~2,~3$, respectively.
 The capital indices, meanwhile, are left unaltered. $M$ denotes the mass
 and $a$ the spin parameter of the central Kerr black hole.

 The tetrad we use has been provided in \cite{Barausse09} and reads
 \begin{eqnarray}
  \tilde{e}^T_\mu &=& \delta^t_\mu \sqrt{\frac{\Delta\Sigma}{\Lambda}}~~,\nonumber\\
  \tilde{e}^1_\mu &=& \delta^r_\mu \sqrt{\frac{\Sigma}{\Delta}}~~,\nonumber\\
  \tilde{e}^2_\mu &=& \delta^\theta_\mu \sqrt{\Sigma}~~,\nonumber\\
  \tilde{e}^3_\mu &=& -\delta^t_\mu 
  \frac{2 a M r \sin \theta}{\sqrt{\Lambda\Sigma}}
  +\delta^\phi_\mu \sin{\theta} \sqrt{\frac{\Lambda}{\Sigma}}~~, \label{eq:tetrad}
 \end{eqnarray}
 while the inverse one reads
 \begin{eqnarray}
  \tilde{e}_T^\mu &=& \delta_t^\mu \sqrt{\frac{\Lambda}{\Delta\Sigma}}
  + \delta_\phi^\mu  \frac{2 a M r}{\sqrt{\Delta\Lambda\Sigma}}~~,\nonumber\\
  \tilde{e}_1^\mu &=& \delta_r^\mu \sqrt{\frac{\Delta}{\Sigma}}~~,\nonumber\\
  \tilde{e}_2^\mu &=& \delta_\theta^\mu \frac{1}{\sqrt{\Sigma}}~~,\nonumber\\
  \tilde{e}_3^\mu &=& \delta_\phi^\mu \frac{1}{\sin{\theta}}
  \sqrt{\frac{\Sigma}{\Lambda}}~~.\label{eq:invtetrad}
 \end{eqnarray}

 By calculating all the quantities mentioned in Sec.~\ref{sec:HamSP}, we finally 
 obtain the coefficients $H^C$ (Eq.~(\ref{eq:HamSPc})) as

\begin{widetext}
 \begin{eqnarray}
  H^1 &=& -\frac{\sqrt{\Delta}\cos \theta}{\sqrt{Q}(1+\sqrt{Q})\Lambda^2\sqrt{\Sigma}\sin^2 \theta}
 [(1+\sqrt{Q})(\Delta \Sigma^2+2~M~r\varpi^4)+\sqrt{Q} 2 a^2 M r \varpi^2 \sin^2\theta]\frac{P_\phi}{\mu} \nonumber \\
  &+& \frac{a M (2 r^2 \Sigma+\varpi^2 \rho^2) \sin \theta~\Delta}
 {\sqrt{Q}(1+\sqrt{Q})\Lambda^{3/2} \Sigma^{2}}\frac{P_r P_\theta }{\mu^2}
 +\frac{2 a^3 M r \cos \theta \sin^2 \theta~\Delta}
          {\sqrt{Q}(1+\sqrt{Q})\Lambda^{3/2} \Sigma }
 \left(1+\sqrt{Q}+\frac{2 \Sigma}{\Lambda \sin^2\theta}\frac{P_\phi^2}{\mu^2}
  +\frac{\Delta}{\Sigma}\frac{P_r^2}{\mu^2}\right)~~, \nonumber \\
   H^2 &=& \frac{\Delta(1+\sqrt{Q})(r\Sigma^2-a^2 M \rho^2 \sin^2 \theta)
   -M \sqrt{Q}(\rho^2 \varpi^4-4 a^2 M r^3 \sin^2\theta)}
  {\sqrt{Q}(1+\sqrt{Q})\Lambda^2\sqrt{\Sigma}\sin\theta}~\frac{P_\phi}{\mu}
  +\frac{2 a^3 M r \cos \theta \sin^2 \theta~\Delta^{3/2}}
          {\sqrt{Q}(1+\sqrt{Q})\Lambda^{3/2} \Sigma^2 }\frac{P_r P_\theta}{\mu^2}\nonumber\\
  &+& \frac{a M (2 r^2 \Sigma+\varpi^2 \rho^2) \sin \theta~\sqrt{\Delta}}
 {\sqrt{Q}(1+\sqrt{Q})\Lambda^{3/2} \Sigma}
\left(1+\sqrt{Q}+\frac{2 \Sigma}{\Lambda \sin^2\theta}\frac{P_\phi^2}{\mu^2}
  +\frac{1}{\Sigma}\frac{P_\theta^2}{\mu^2}\right)~~, \nonumber \\
   H^3 &=& -\frac{a^2 \Delta \cos \theta \sin \theta}{\sqrt{Q}(1+\sqrt{Q})(\Lambda \Sigma)^{3/2}}
   (\Lambda+\sqrt{Q} \Delta \Sigma)\frac{P_r}{\mu}
  -\frac{r \Lambda \Delta+\varpi^2 \Sigma \sqrt{Q}(r \Delta-M(r^2-a^2))}
   {\sqrt{Q}(1+\sqrt{Q})(\Lambda \Sigma)^{3/2}}~\frac{P_\theta}{\mu}\nonumber\\
   &-& \frac{a M \sqrt{\Delta}}{\mu^2\sqrt{Q}(1+\sqrt{Q})\Lambda^2 \Sigma }
[2 a^2 r \Delta \sin \theta \cos \theta~P_r+(2 r^2 \Sigma+\varpi^2 \rho^2) P_\theta] P_\phi  ~~,
 \end{eqnarray}
 \end{widetext}
 where
 \begin{eqnarray}
  Q &=& 1+\frac{\gamma^{ij}}{\mu^2} P_i P_j  \\
  ~ &=& 1+\mu^{-2}\left(\frac{\Delta}{\Sigma} P^2_r+\frac{1}{\Sigma} P^2_\theta
        + \frac{\Sigma}{\Lambda \sin^2 \theta} P^2_\phi \right)~~,\nonumber
 \end{eqnarray}
 and
 \begin{equation}
  \rho^2 = r^2-a^2 \cos^2\theta~~.
 \end{equation}
 For a full and detailed presentation of the derivation of $H^C$, we refer the
 reader to \cite{Barausse09}.

 It is worth mentioning here that, contrary to the T~SSC, the NW~SSC
 (Eq.~\eqref{eq:NWTimeLV}) does not uniquely define the reference worldline. As
 already noted in the introduction the choice of the center of mass, i.e., the
 reference worldline, is observer dependent. When T~SSC is applied the zero
 3-momentum observer is chosen. However, when the NW~SSC is used there is no unique
 choice because the observer and therewith the reference worldline depends on the
 tetrad. We have fixed our tetrad in Eqs.~\eqref{eq:tetrad},~\eqref{eq:invtetrad}. 
 In the following we only consider the evolution of the orbit corresponding to this
 observer so that we do not have to worry about transforming the dynamical
 properties of the system to another reference frame.

\section{Comparison of Tulczyjew and Newton-Wigner SSC}\label{sec:TNWCom}

\subsection{Preliminaries} \label{subsec:TNWprel}

 When simulating the MP equations we in fact have to solve the initial value problem
 \begin{align}\label{eqn-initial-value-problem-MP}
 \begin{cases}
  &\frac{\mathrm d~x^\mu}{\mathrm d \tau} =v^\mu~~,\\
  &\frac{\mathrm d~p^\mu}{\mathrm d \tau}=-\frac{1}{2}~{R^\mu}_{\nu\kappa\lambda}
  v^\nu S^{\kappa\lambda}-\Gamma^\mu_{\nu\kappa}v^{\nu}p^{\kappa}~~,\\
  &\frac{\mathrm d~S^{\mu\nu}}{\mathrm d \tau}=p^\mu~v^\nu-v^\mu~p^\nu+
  \Gamma^\mu_{\kappa\lambda}S^{\nu\kappa}v^{\lambda}-\Gamma^{\nu}_{\kappa\lambda}S^{\mu\kappa}v^{\lambda}~~,\\ 
  &x^\mu(\tau=0)=x^\mu_0~~,\\
  &p^\mu(\tau=0)=p^\mu_0~~,\\
  &S^{\mu\nu}(\tau=0)=S^{\mu\nu}_0~~.
  \end{cases}
 \end{align}
 As a first step, we have to provide initial conditions which comply with the
 constraints mentioned earlier (Sec.~\ref{sec:MPeqs}).

 In order to find these appropriate initial conditions, we follow the approach
 given  in \cite{Hartl03}, which implies that instead of the spin tensor
 $S^{\mu\nu}$ we use the vector $S^\mu$ for the initial setup. Without loss of
 generality, we set $t=\phi=0$ and provide initial values for $r,~\theta,~p^r$ 
 as well as for the two spin components $S^r$ and $S^\theta$. The other initial
 conditions, namely $p^t$, $p^\theta$, $p^\phi$, $S^t$, and $S^\phi$, are then
 fixed by the constraints. In the case of the T~SSC, those constraints are
 \begin{align}
 E &= -p_t-\frac{1}{2 \mu} g_{t\mu,\nu}\eta^{\mu\nu\gamma\delta} S_{\gamma} p_\delta~~,\label{eqn-initial-E}\\
 J_z &= p_\phi+\frac{1}{2 \mu} g_{\phi\mu,\nu}\eta^{\mu\nu\gamma\delta} S_{\gamma} p_\delta~~,\label{eqn-initial-J_z}\\
 \mu^2 &= -g^{\mu\nu}p_\mu p_\nu~~,\label{eqn-initial-mu^2}\\
 S^2 &= g^{\mu\nu}S_\mu S_\nu~~,\label{eqn-initial-S^2}\\
 0 &= g^{\mu\nu} S_\mu p_\nu~~,\label{eqn-initial-S-ortho-p}
 \end{align}
 where we have substituted Eq.~(\ref{eq:T4VSin}) into the constants of
 motion~(\ref{eqn-constant-of-E}),~(\ref{eqn-constant-of-J_z}), and lowered the indices
 wherever needed. Thus, we specify an orbit by providing values for $E$, $J_z$,
 $S^2$, and $\mu^2$. We then solve the
 system~\eqref{eqn-initial-E}-\eqref{eqn-initial-S-ortho-p} for
 $p_t$, $p_\theta$, $p_\phi$, $S_t$, and $S_\phi$ with the help of the 
 Newton-Raphson method.
 
 For comparing the effect of different SSCs in the evolution of MP, we need to find
 initial conditions for the NW~SSC which are similar to the T~SSC case. Hence, we
 parametrize the orbits by providing the same initial set of values for $r$, 
 $\theta$,  $p^r$, $S^r$, $S^\theta$, $E$, $J_z$, $S^2$ and $\mu^2$. The set of
 constraints for the NW~SSC is similar to the one for the T~SSC
 (Eqs.~\eqref{eqn-initial-E}-\eqref{eqn-initial-S-ortho-p}). The
 constraints~\eqref{eqn-initial-E}-\eqref{eqn-initial-J_z} remain unaltered. We use
 Eq.~\eqref{eqn-initial-mu^2}, and Eq.~\eqref{eq:NWSpinMeas4V} instead of
 Eq.~\eqref{eqn-initial-S^2} for the initial setup, even though, in the case of the
 NW SCC, neither the spin $S^2$ nor the rest mass $\mu$ is preserved anymore.
 Finally, we replace constraint~\eqref{eqn-initial-S-ortho-p} by 
 \begin{align}
  g^{\mu\nu}S_\mu \omega_\nu&=0~~.\nonumber
\end{align}
 When solving the resulting system for $p_t$, $p_\theta$, $p_\phi$, $S_t$, and
 $S_\phi$ for the same provided $r$, $\theta$,  $p^r$, $S^r$, $S^\theta$, $E$,
 $J_z$, $S^2$ and $\mu^2$ as in the T case, we get what we referred to as similar
 initial conditions above. At last, by raising indices of the momenta and going
 from spin vectors to tensors with the help of the
 transformations~\eqref{eq:T4VSin}~and~\eqref{eq:NW4VSin}, respectively, we get
 suitable data to start the computation with. The orbits are evolved through the
 Eqs.~\eqref{eq:MPmomenta},~\eqref{eq:MPspin}. A more detailed discussion about the
 techniques we have applied to evolve the MP equations is provided in
 Appendix~\ref{sec:NumIntMP}.

 The timelike vector $\omega_\nu$ in the NW~SSC~\eqref{eq:NW} is given by
 Eq.~\eqref{eq:NWTimeLV}, where the $\tilde{e}^T_\nu$ is the top equation from
 the set~\eqref{eq:tetrad}. By adapting the convention that times and lengths are
 measured in terms of $M$, we set $M=1$ throughout the paper.

 Before we proceed with the numerical results, we want to discuss the initial setup
 for our evaluations in this section. We have chosen the orbits to start from the
 same point in the configuration space, i.e., both worldlines at $\tau=0$ lie at
 the same spacetime point. This means that both of the different corresponding
 observers see the center of the mass lying at the same place, even if the SSCs are
 different. This is not the usual way this subject is treated. In \cite{Kyrian07},
 for example, the discussion about the transition between two different SSCs is
 based on the center of the mass worldline displacement. The latter approach would
 not be appropriate for our treatment, because apart from the shift in the value of
 the spin tensor, the initial point in the configuration space should be shifted as
 well \cite{Kyrian07}. In our treatment we want to change the order of magnitude of 
 the spin while keeping the initial conditions as similar as possible during the 
 scaling, in order to observe how the two different SSCs converge as the geodesic
 limit is approached. In other words we do not attempt to have initial conditions
 which would obey the transition between different SSCs for one particle, but rather
 conditions which represent similar orbits for two different SSCs.

 \subsection{Comparison for large spin}
 \begin{figure*} [htp]
  \centerline{\includegraphics[width=0.3 \textwidth] {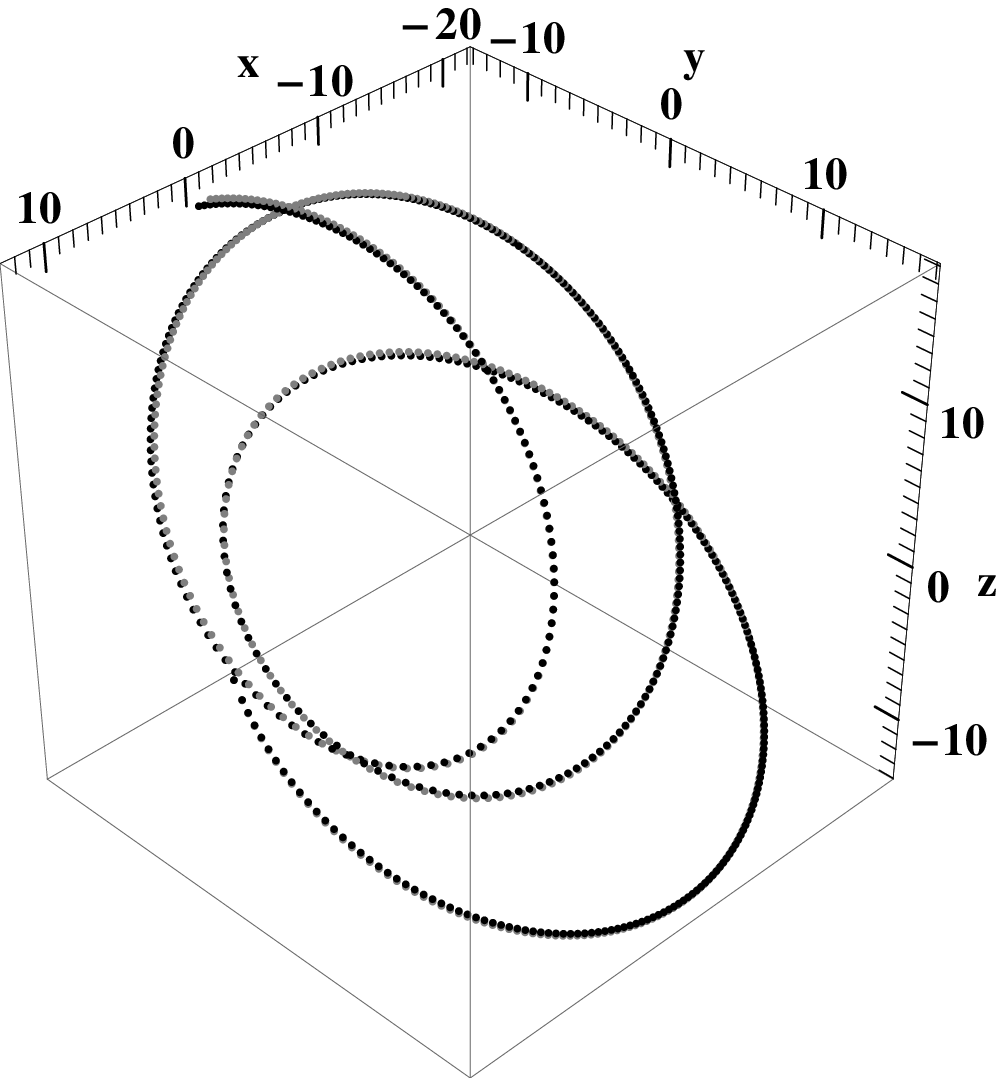}
  \includegraphics[width=0.3 \textwidth] {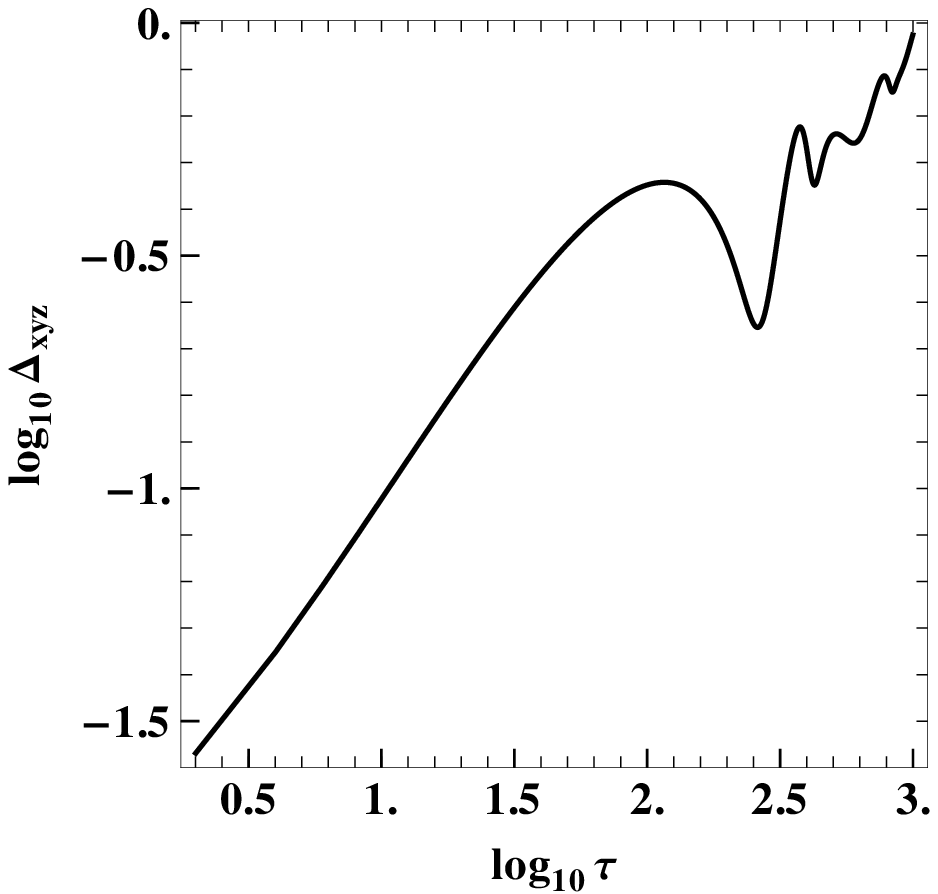}
 \includegraphics[width=0.3 \textwidth]{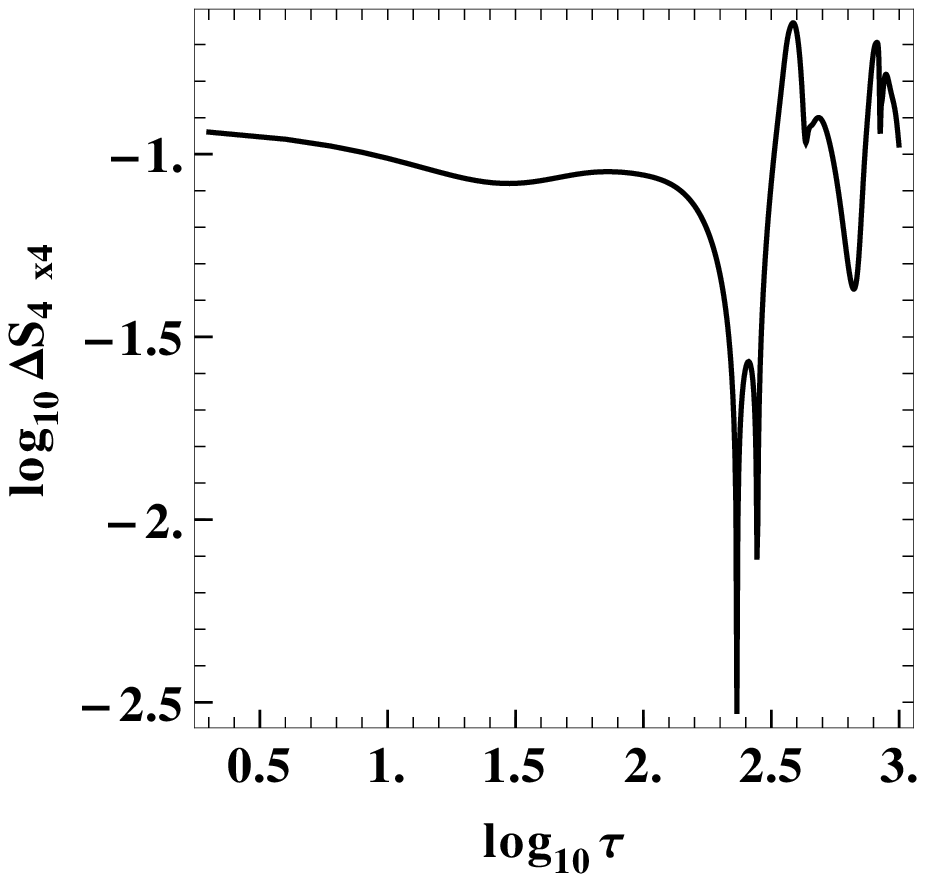}
 }
 \caption{ The left panel shows a MP orbit with T~SSC (black dots) and a MP orbit
 with NW~SSC (gray dots) in the configuration space $x,~y,~z$ (Cartesian coordinates).
 The common parameters for these orbits are $a=0.5$, $r=11.7$, $\theta=\pi/2$, 
 $p^r=0.1$, $S=1$, $S^r=0.1~S$, $S^\theta=0.01~S$, $E=0.97$, $J_z=3$, and $\mu=1$. 
 The central panel shows the logarithm of the Euclidean distance in the
 configuration space between these two orbits as a function of the proper time.
 The right panel shows the logarithm of the difference $\Delta S_{4\textrm{x}4}$
 between the spin tensors of these two orbits as a function of the proper time.
 }
 \label{Fig:ConSpCarta05m0}
 \end{figure*} 

 In our first example, the parameters read $a=0.5$, $r=11.7$, $\theta=\pi/2$,
 $p^r=0.1$, $S=1$, $S^r=0.1~S$, $S^\theta=0.01~S$, $E=0.97$, $J_z=3$, and $\mu=1$.
 The left panel of Fig.~\ref{Fig:ConSpCarta05m0} shows how the two MP orbits with T
 SSC (black) and NW~SSC (gray) evolve in the configuration space where the
 Cartesian coordinates
 \begin{eqnarray}
  x &=& r \cos\phi \sin\theta~~,\nonumber \\
  y &=& r \sin\phi \sin\theta~~,\nonumber \\
  z &=& r \cos\theta~~,\label{eq:CarCoor}
 \end{eqnarray} 
 are employed. 

 The divergence between the two orbits is barely visible in the left panel, but if
 we take the Euclidean norm
 \begin{equation}
  \Delta_{xyz}=\sqrt{(x_{T}-x_{NW})^2+(y_{T}-y_{NW})^2+(z_{T}-z_{NW})^2}~~,
 \label{eq:EuclNorCarCoor} 
 \end{equation}
 we see that at the end of our run, the separation between the two orbits is of
 the order one (central panel of Fig.~\ref{Fig:ConSpCarta05m0}), while the
 radial distance from the central black hole is of the order ten (left panel of
 Fig.~\ref{Fig:ConSpCarta05m0}). Even if the M\"{o}ller radius is not an
 appropriate tool for our setup (see the discussion at the last paragraph of
 Sec.~\ref{subsec:TNWprel}), it is worthy to note that the two orbits lie inside
 a M\"{o}ller radius ($S/\mu=1$) for $\tau=10^3$, even if their distance  will grow
 out of this radius later on. This divergence in the orbit evolution follows the 
 discrepancy in the spin space. To illustrate this, the norm of the difference
 between the spin tensor $S^{\mu\nu}_{T}$ of the T~SSC and the spin tensor
 $S^{\mu\nu}_{T}$ of the NW~SSC,
 \begin{equation}
  \Delta S_{4\textrm{x}4}=\sqrt{\left|g_{\mu\nu}g_{\kappa\lambda} (S^{\nu\kappa}_{T}-S^{\nu\kappa}_{NW})
   (S^{\mu\lambda}_{T}-S^{\mu\lambda}_{NW})\right|}~~,
 \label{eq:DS4x4}    
 \end{equation}
 is displayed in the right panel of Fig.~\ref{Fig:ConSpCarta05m0}.
 $\Delta S_{4\textrm{x}4}$ is one tenth of the spin measure right from the 
 beginning, and stays at this level during the evolution. Thus, from an orbital
 dynamic point of view when the spin of the test particle is of order $S=1$, the
 choice of different SSCs leads to orbital evolutions which diverge significantly with
 time.

 One thing that has to be discussed before we proceed is the meaning of a 'common'
 proper time, when two orbits with different SSCs are compared. Each SSC defines its
 own center of reference, which implies that with each SSC the proper time that is
 measured along the above orbits is different, even if the orbits start with
 similar initial conditions. Another issue that arises here is how we can measure
 the distance between two `nearby' orbits in a curved spacetime. Above, we use the
 Euclidean norm, however the spacetime is not Euclidean. The same issues arise when
 geodesic chaos is studied in curved spacetimes (see, e.g., \cite{chaos}). One of
 the suggestions in the aforementioned field is to use the two nearby orbits 
 technique, i.e., to evolve two orbits with similar initial conditions and measure
 their distance when they reach the same proper time. This is in few words the
 approach we adapt in our study for the time issue. For the issue of the distance
 in the configuration space between the two orbits, we have chosen to employ
 the Euclidean metric. We could employ the local $g_{\mu\nu}$ metric as well,
 even if the orbits depart from each other significantly (middle panel of
 Fig.~\ref{Fig:ConSpCartHa05m0}). However, for the evolution times $\tau=10^3$ the
 results coming from both approaches are almost identical, and therefore we went
 for the the simplest metric, which is the Euclidean.

 \subsection{Comparison for very small spin}
 \begin{figure} [htp]
  \centerline{
  \includegraphics[width=0.25 \textwidth] {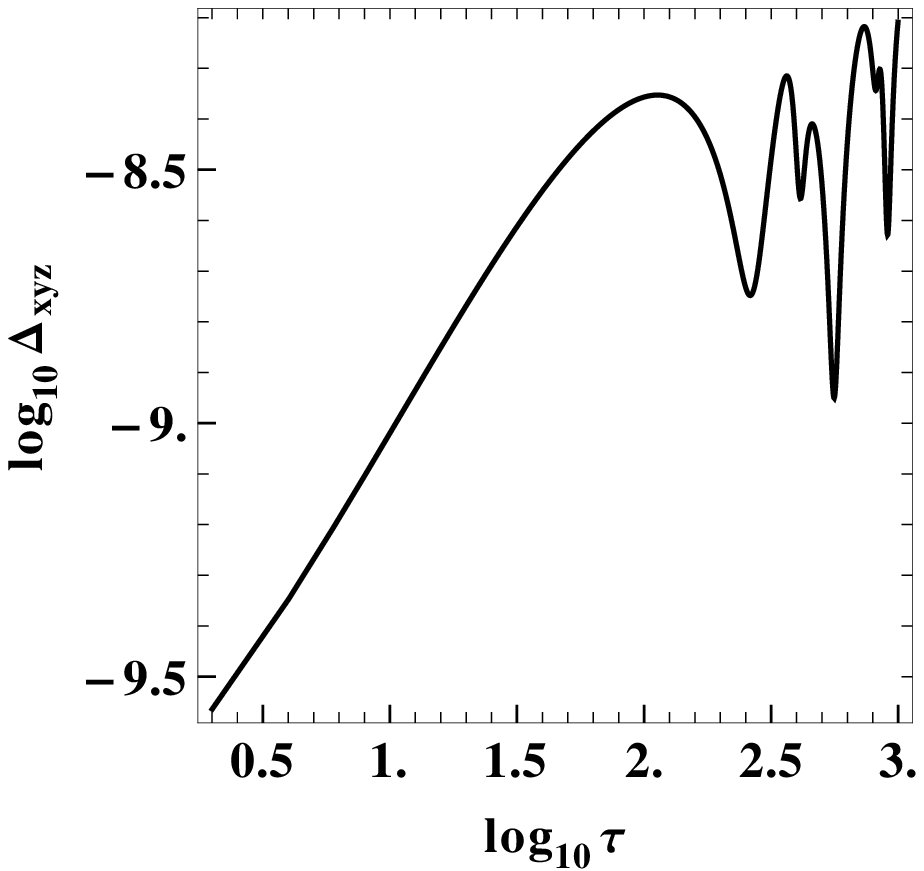}
 \includegraphics[width=0.25 \textwidth]{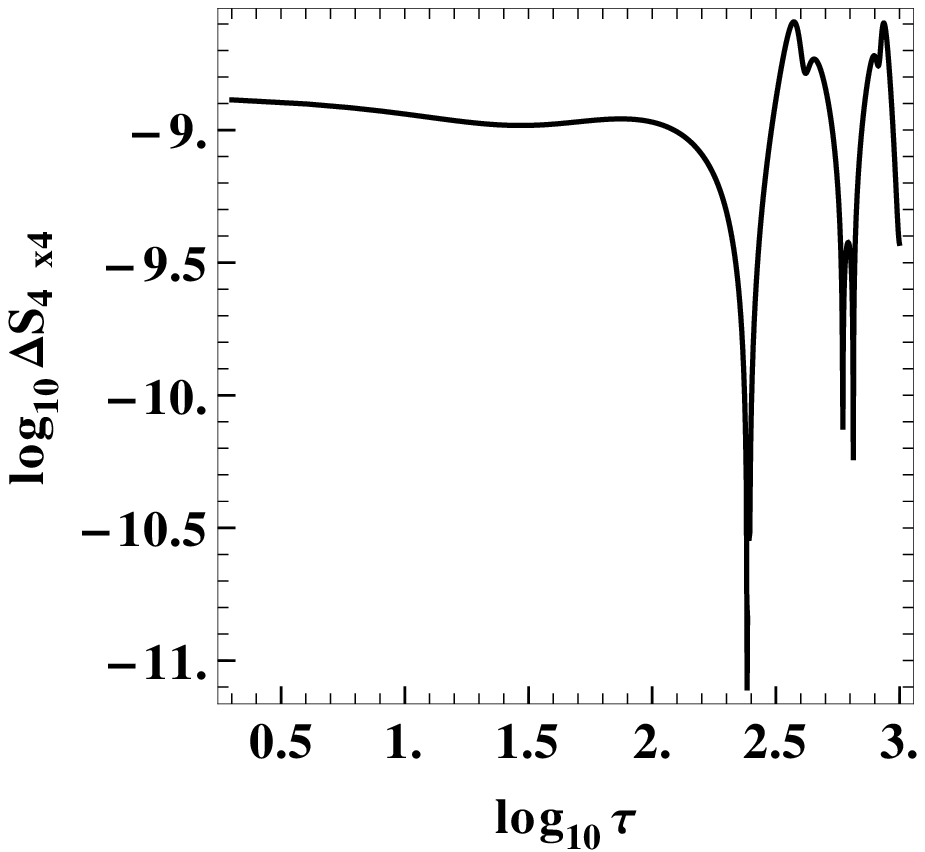}
 }
 \caption{ The left panel shows the logarithm of the Euclidean distance in the
 configuration space between a MP orbit with T~SSC and a MP orbit with NW~SSC as a 
 function of the proper time. The common parameters for these orbits read $a=0.5$,
 $r=11.7$, $\theta=\pi/2$, $p^r=0.1$, $S=10^{-8}$, $S^r=0.1~S$, $S^\theta=0.01~S$,
 $E=0.97$, $J_z=3$, and $\mu=1$. The right panel shows the logarithm of the
 difference $\Delta S_{4\textrm{x}4}$ between the spin tensors of these two orbits
 as a function of the proper time.
 }
 \label{Fig:ConSpCarta05m8}
 \end{figure}

 Since we mentioned the geodesic orbits, we approach this limit by setting the 
 measure of the spin in our initial conditions to $S=10^{-8}$. All the other
 parameters are the same as in Fig.~\ref{Fig:ConSpCarta05m0}. For this geodesic-like
 setup the orbits in the configuration space resemble the orbits shown in the left
 panel of Fig.~\ref{Fig:ConSpCarta05m0}. However, the left panel of
 Fig.~\ref{Fig:ConSpCarta05m8} shows that the distance between the two orbits has
 dropped significantly, about $8$ orders of magnitude. This drop is
 anticipated since we tend to the geodesic limit and the spin contribution is 
 expected to be smaller. However, the level of the divergence in the configuration 
 space (left panel of Fig.~\ref{Fig:ConSpCarta05m8}) is again defined by the
 magnitude of the spin difference $\Delta S_{4\textrm{x}4}$  (right panel of
 Fig.~\ref{Fig:ConSpCarta05m8}). Namely, even though the initial conditions in
 the configuration space are identical, i.e., $\Delta_{xyz}=0$, those of the 
 spin components are not, i.e.,  $\Delta S_{4\textrm{x}4}\approx 10^{-9}$, and this
 initial divergence in the spin space is passed on to the configuration space.

 \subsection{Constants of motion}
 \begin{figure} [htp]
  \centerline{
  \includegraphics[width=0.25 \textwidth] {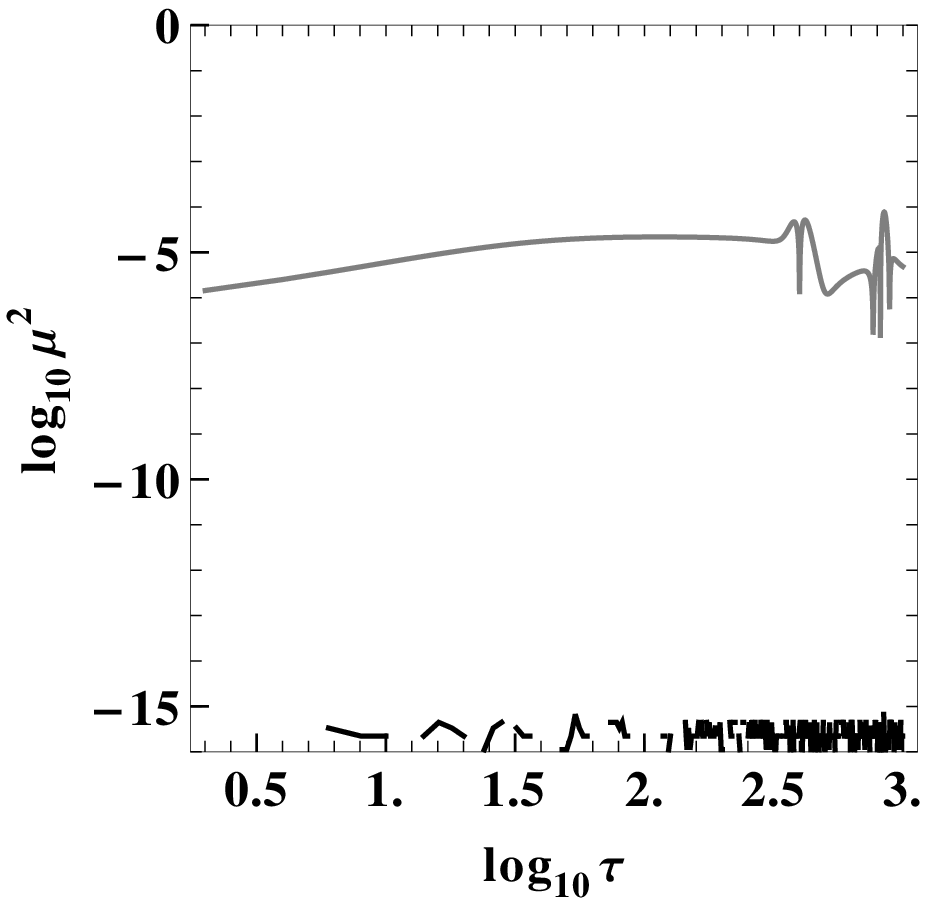}
 \includegraphics[width=0.25 \textwidth]{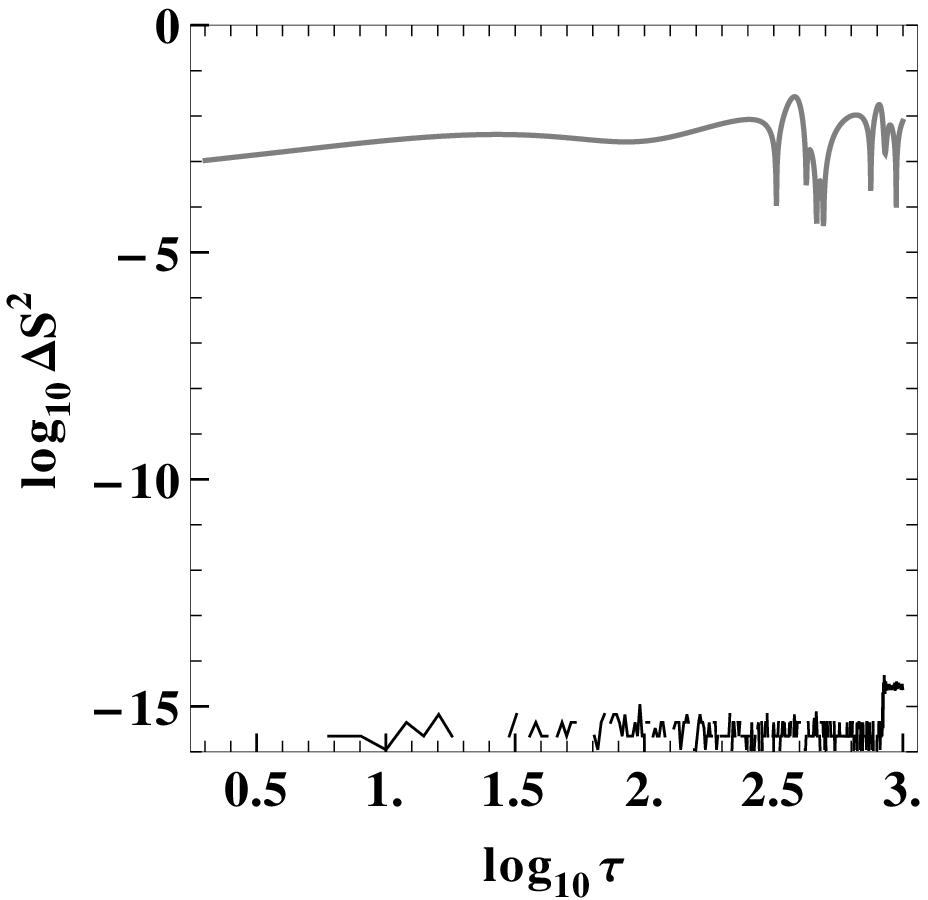}
 }
  \centerline{
  \includegraphics[width=0.25 \textwidth] {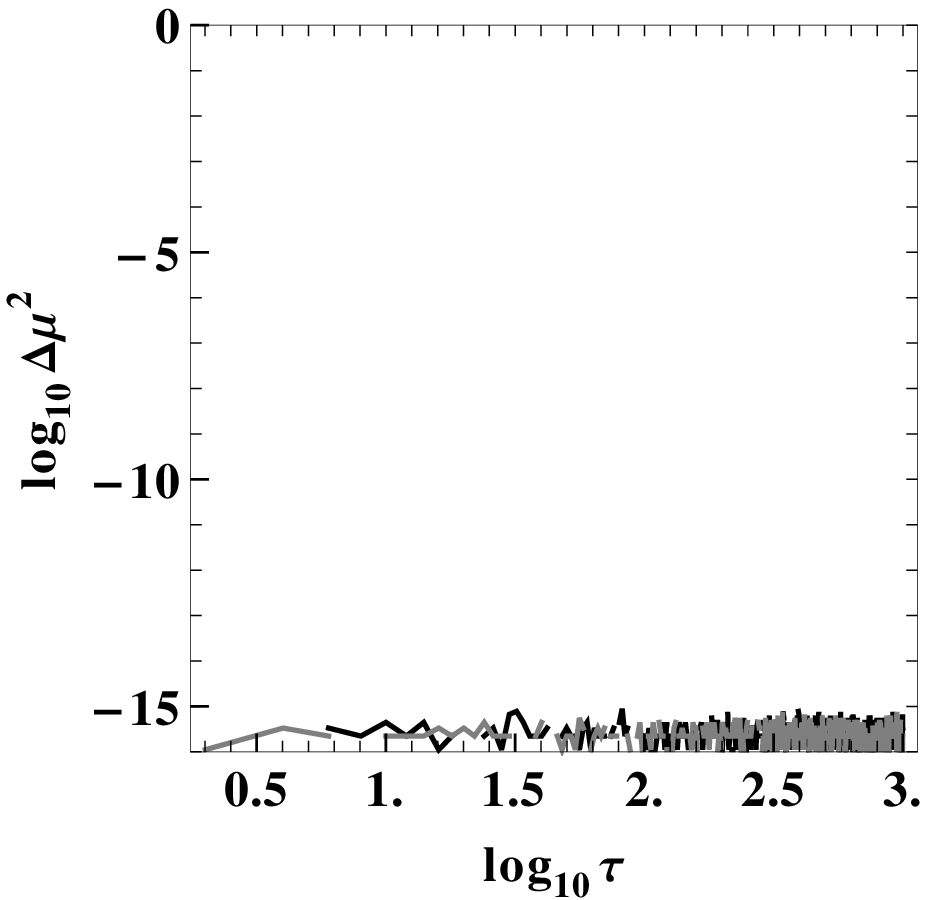}
 \includegraphics[width=0.25 \textwidth]{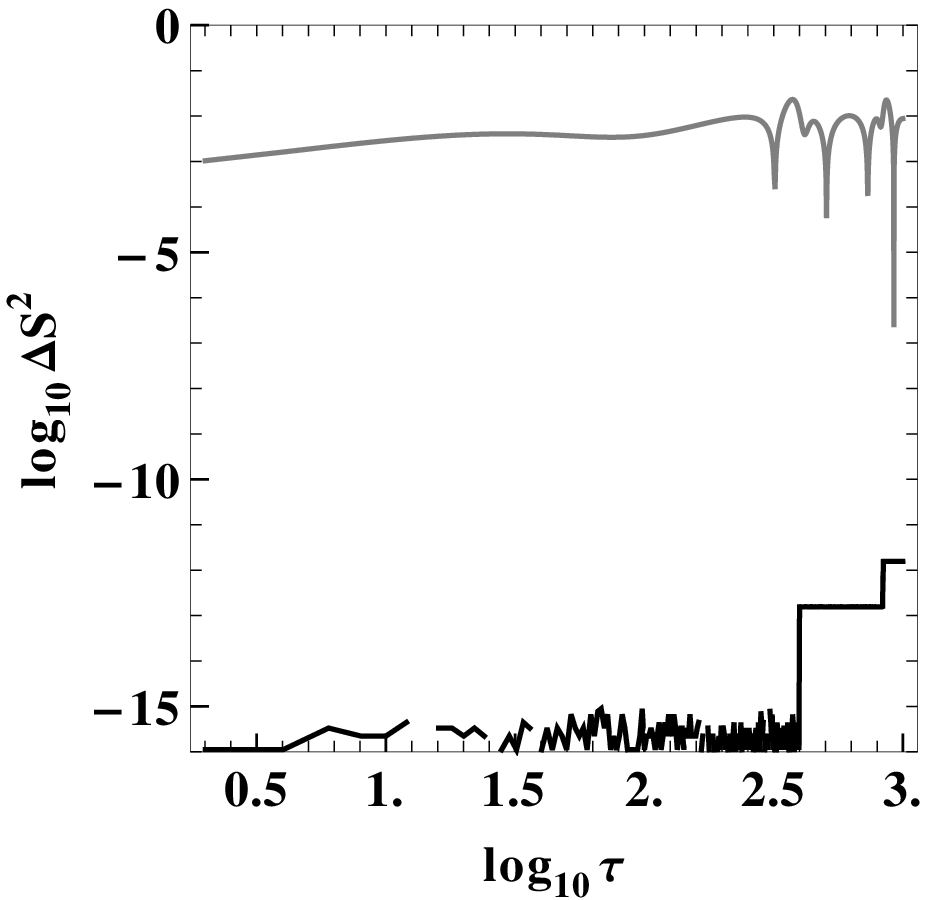}
 }
 \caption{ The top row of panels corresponds to the orbits of
 Fig.~\ref{Fig:ConSpCarta05m0}, while the bottom row of panels corresponds to the
 orbits of Fig.~\ref{Fig:ConSpCarta05m8}. The black lines represent the evolution
 of the MP equations with T~SSC, while the gray lines represent the NW~SSC.
 The left column of panels shows the relative error in the preservation of the 
 four-momentum, while the right depicts the preservation of the spin.  
 }
 \label{Fig:TNWDmu2DS2a05m0m8}
 \end{figure} 
 
 We now turn our attention to the conservation of the four-momentum (rest mass
 $\mu$) and of the spin $S$. In order to check whether these quantities are
 preserved, we use the relative error of the four-momentum 
 \begin{equation}
  \Delta \mu^2=\left|1-\frac{\mu^2(\tau)}{\mu^2(0)}\right|~~,\label{eq:RelEr4m}
 \end{equation}
 and the relative error of the spin $S^2$
 \begin{equation}
  \Delta S^2=\left|1-\frac{S^2(\tau)}{S^2(0)}\right|~~,\label{eq:RelErS2}  
 \end{equation}
 where $\mu^2(\tau)$, and $S^2(\tau)$ are calculated at time $\tau$.

 We see from Fig.~\ref{Fig:TNWDmu2DS2a05m0m8} that both the rest mass $\mu^2$
 and the spin are conserved for the T~SSC (black lines) as was expected
 (see Sec.~\ref{sec:MPeqs}). On the other hand, in the case of the NW~SSC (gray lines)
 the four-momentum scales with the magnitude of the spin $S$, while the square of
 the spin itself stays at the same level indifferently from the spin's magnitude.
 This scaling in the conservation of the mass is anticipated because, as
 $S\rightarrow 0$, the evolution of the MP equations approaches that of the geodesic
 motion. 
 \begin{figure} [htp]
  \centerline{
  \includegraphics[width=0.3 \textwidth] {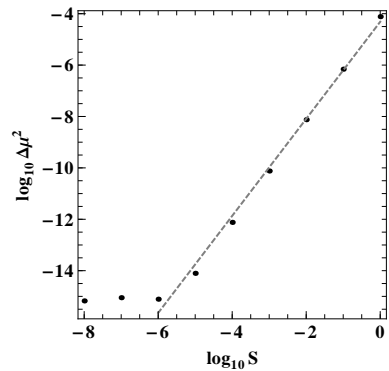}
 }
 \caption{The relative error of the four-momentum $\Delta \mu^2$ as a function of
 the  spin measure $S$ for the NW~SSC. The black dots correspond to the maximum
 values of $\Delta \mu^2$ during the evolution for each $S$. The dashed line is a
 linear fit of the form  $\log_{10}{\Delta\mu^2}=a\log_{10}{S}+b$  for 
 data with $S>10^{-6}$,
 where $a=1.995\pm0.004,~b=-4.136\pm0.013$.
 }
 \label{Fig:MaSpSca05}
 \end{figure} 

 In order to better illustrate the above mentioned scaling, we run several
 simulations with initial setups similar to the one of Fig.~\ref{Fig:ConSpCarta05m0}
 where we only change the measure of the spin, $S$. For every simulation, we plot
 the maximum value of $\Delta\mu^2$ along the trajectory against the initial
 spin measure (Fig.~\ref{Fig:MaSpSca05}). The resulting plot shows that, as 
 we decrease $S$, the four-momentum for the NW~SSC tends to be conserved up to the 
 computational accuracy. There are two effects that shape this figure. One is 
 the theoretical scaling of $\Delta \mu^2$ as a function of $S$ and the other
 is the finite computational accuracy. From a linear fit of our data we get for 
 $S>10^{-6}$ (dashed line in Fig.~\ref{Fig:MaSpSca05}) $\Delta~\mu^2~\propto~S^2$. 
 For smaller spins a plateau appears because we reach the computational accuracy
 (in our runs we use double precision). 
 
 Since for T~SSC the four-momentum is conserved and for the NW~SSC the 
 $\sqrt{\Delta~\mu^2}$ scales linearly with the spin, this scaling can be interpreted
 as the rate by which the two different SSCs converge to each other. Changing the value
 of the spin $a$ of the central black hole does not alter qualitatively the results of
 our numerical comparison. 

\section{Numerical comparison of the MP equations with the corresponding 
Hamiltonian equations} \label{sec:MPHamCom} 

\subsection{Preliminaries}

  \begin{figure*} [htp]
  \centerline{\includegraphics[width=0.3 \textwidth] {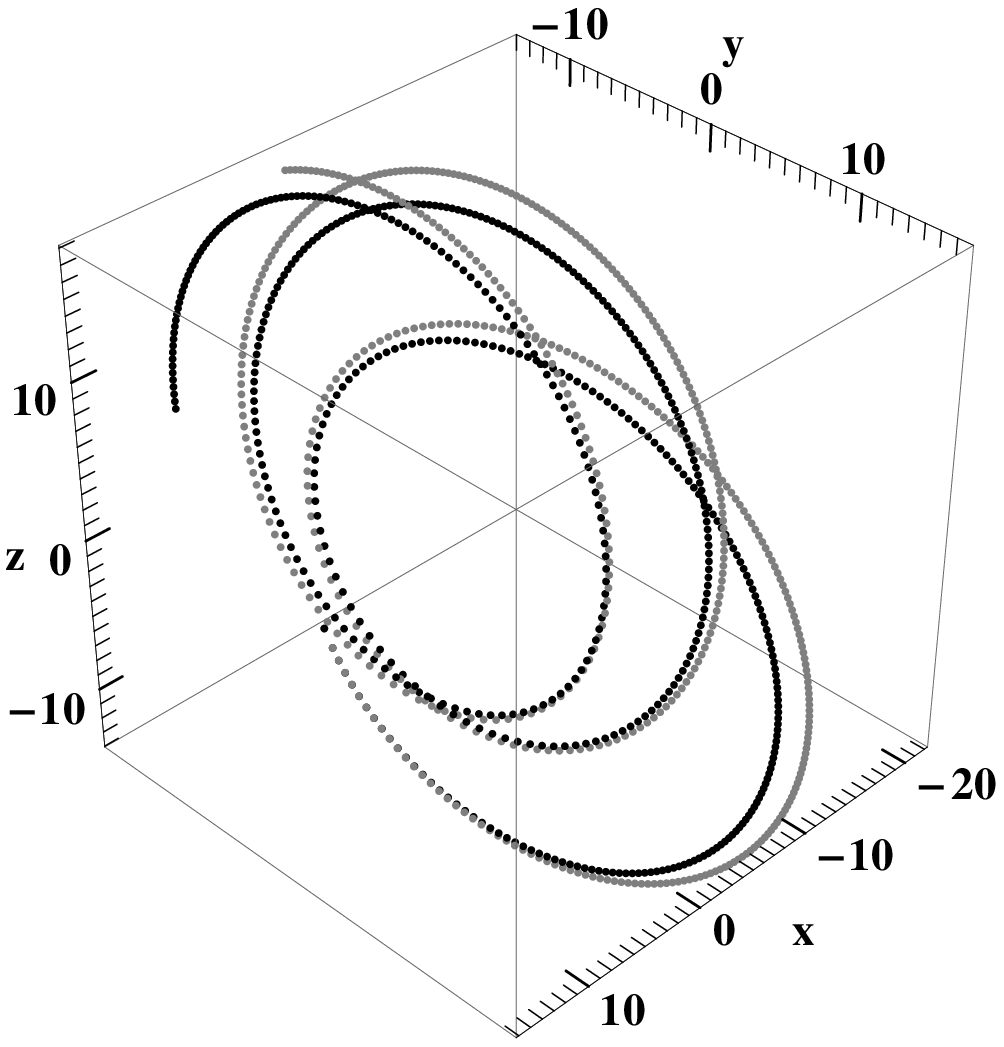}
  \includegraphics[width=0.3 \textwidth] {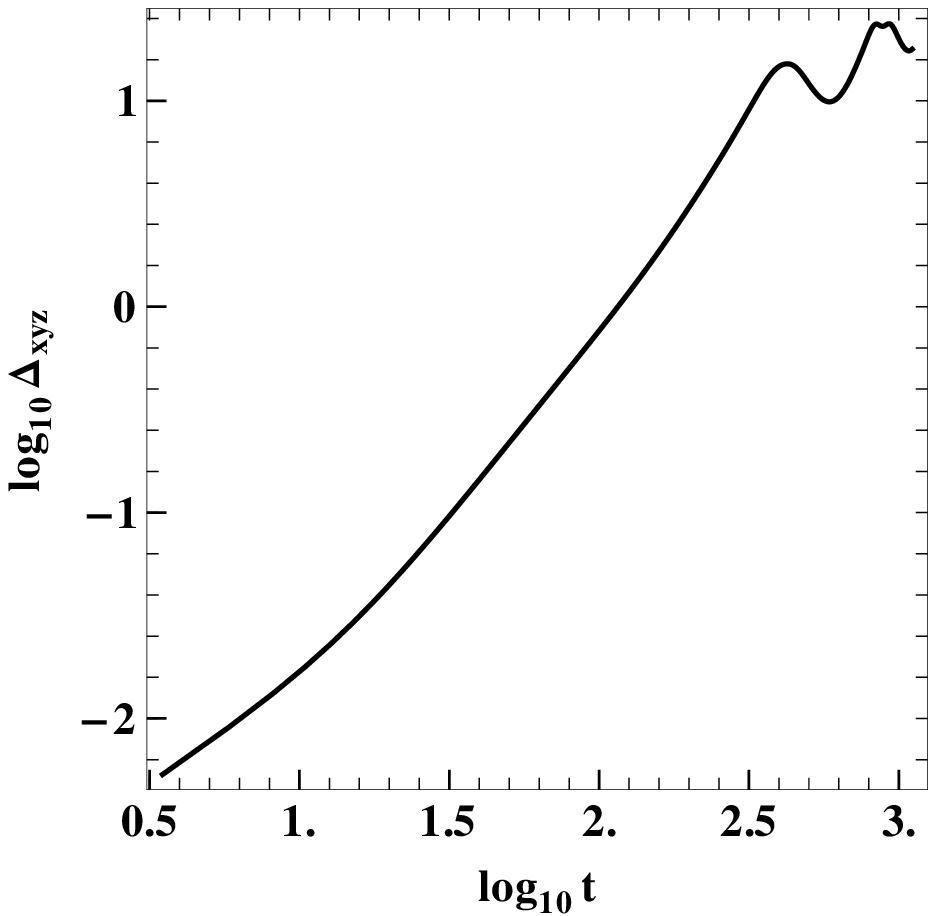}
 \includegraphics[width=0.3 \textwidth]{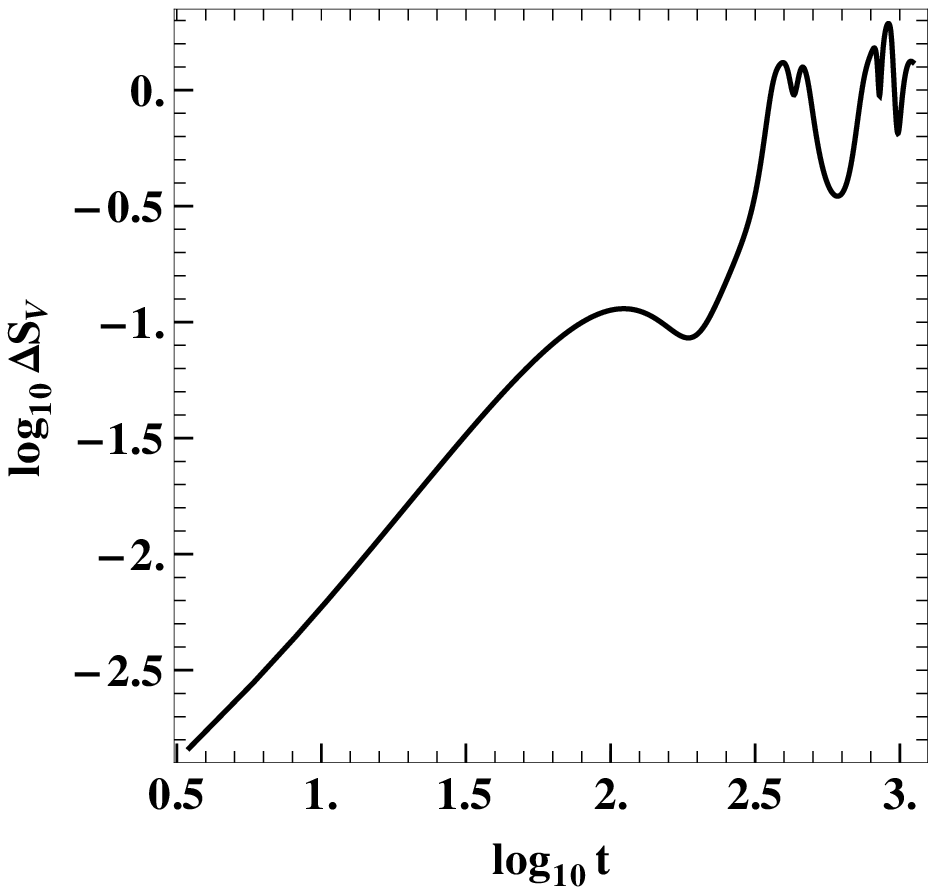}
 }
 \caption{ The left panel shows how the orbit
 evolves through the MP equations (gray dots) and through the Hamilton's equations
 (black dots) in the configuration space  $x,~y,~z$, when we use the initial 
 conditions given in Fig.~\ref{Fig:ConSpCarta05m0}. The central panel shows the
 logarithm of the Euclidean distance in the configuration space between these two
 orbits as a function of the coordinate time. The right panel shows the logarithm 
 of the Euclidean norm of the difference between the spin vectors of these two 
 orbits as a function of the coordinate time.
 }
 \label{Fig:ConSpCartHa05m0}
 \end{figure*} 

 Since the MP equations are a pole-dipole approximation, multipoles of higher order
 than the spin dipole are already neglected. However, we can simplify the problem 
 further by assuming that the physically relevant values for the particle spin 
 are small and the terms quadratic in the spin correspond to the quadrupole
 contribution. Thus, a Hamiltonian which is accurate up to linear order of the
 spin should yield satisfactory results. This is the main idea on which the
 construction of such a Hamiltonian formalism for NW~SSC in~\cite{Barausse09}
 was based.

 According to this formalism (see the brief description in Sec.~\ref{sec:HamSP}), 
 the evolution parameter is not the proper time like in the case of
 Sec.~\ref{sec:TNWCom}, but the coordinate time. In order to perform a comparison
 between the MP equations and the corresponding Hamiltonian
 (Sec.~\ref{subsec:HamKerrSP}) equations, we could rewrite our MP code with respect
 to the coordinate time. However, the coordinate times, at which our quantities
 were calculated in the MP simulations, were given as output anyway. With them at
 hand, there is an easier way out. One can evolve the Hamilton's equations of
 motion using constant steps in the coordinate time, and interpolate the solution
 around the required times of output. A more detailed discussion on this topic and
 the numerical methods we have used is given in Appendix~\ref{sec:NumIntHam}.

 Moreover, in order to make the two formalisms comparable, we used the equations
 given in Sec.~\ref{sec:HamSP} to go from the set of variables 
 $\{x^\mu,p^\mu,S^{\mu\nu}\}$ of the MP equations to the set of variables 
 $\{x^i,P^i,S^I\}$ in the Hamiltonian formalism. Note that this holds also for the
 initial conditions, thus both the MP equations and the corresponding Hamilton's
 equations start with exactly the same initial setup.

 Before showing the results of comparisons between the two approaches, we want to
 point out that all simulations using the Hamiltonian equations were much faster 
 than their equivalents based on the MP equations with NW~SSC. More detailed
 information on this can be found in the Appendix Sections.

 \subsection{Comparison for large spin}

 Using the initial conditions for the NW~SSC given in Fig.~\ref{Fig:ConSpCarta05m0},
 we have evolved the orbit by using Hamilton's equations. The motion of the 
 corresponding orbit in the configuration space is shown in the left panel of
 Fig.~\ref{Fig:ConSpCartHa05m0} (black dots) together with the orbit evolved
 through the MP equations (gray dots). Even if the two orbits start with the same 
 initial conditions they depart from each other quite quickly. This is seen
 more clearly in the central panel of Fig.~\ref{Fig:ConSpCartHa05m0}, where the
 Euclidean distance between the two orbits 
 \begin{equation}
  \Delta_{xyz}=\sqrt{(x_{H}-x_{MP})^2+(y_{H}-y_{MP})^2+(z_{H}-z_{MP})^2}~~,
 \label{eq:EuclNorCarCoorH} 
 \end{equation}
 is displayed as a function of the coordinate time. Near the end of the calculation,
 the distance $\Delta_{xyz}$ is almost as large as the radial distance of
 the particle from the central black hole. From the appearance of the left panel of
 Fig.~\ref{Fig:ConSpCartHa05m0} one might wonder whether the divergence between 
 the orbits is a ``synchronization'' issue. However, since both schemes use the same 
 SSC, i.e., the NW~SSC, and since the initial conditions for both schemes are
 exactly the same, i.e., the orbits correspond to the same particle, the proper
 time for both orbits has to tick at the same rate. Thus, it is reasonable to claim
 that this divergence results from the fact that the Hamiltonian is valid up to the
 linear order in the particle spin, and since the spin here is large, i.e., $S=1$,
 such divergence should be expected. Nevertheless, it is impressive that orbits
 corresponding to the same particle evaluated with different schemes, i.e., the MP
 equations and the corresponding Hamiltonian, give a divergence that is of one
 order of magnitude larger than the divergence of the MP equations with different
 SSCs (left panel of Fig.~\ref{Fig:ConSpCarta05m0}). If we took the 
 M\"{o}ller radius as a criterion, for example, then, since the distance between 
 the two orbits exceeds the diameter of the disc of centers of mass, according to
 this criterion, the orbits could not correspond to the same particle. Therefore,
 we can say that the Hamiltonian formalism is not valid for large spin values, 
just as expected.
 
 The spin in the Hamiltonian formalism is given by the projection vector
 (Eq.~\eqref{eq:3ProjSpin}). The Euclidean norm of the difference between the spin
 vector $S^I_{H}$ calculated by Hamilton's equations and the $S^I_{MP}$
 calculated by the MP equations
 \begin{equation}
  \Delta S_\textrm{v}=\sqrt{\sum_{I=1}^3 (S^I_{H}-S^I_{MP})^2}~~,
 \label{eq:EuclNorSpin3V}    
 \end{equation}
 is plotted as a function of the coordinate time in the right panel of
 Fig.~\ref{Fig:ConSpCartHa05m0}. This plot shows that the difference is quite
 high, even if the spin values are identical at first.

 \subsection{Comparison for very small spin}
 \begin{figure} [htp]
  \centerline{
  \includegraphics[width=0.25 \textwidth] {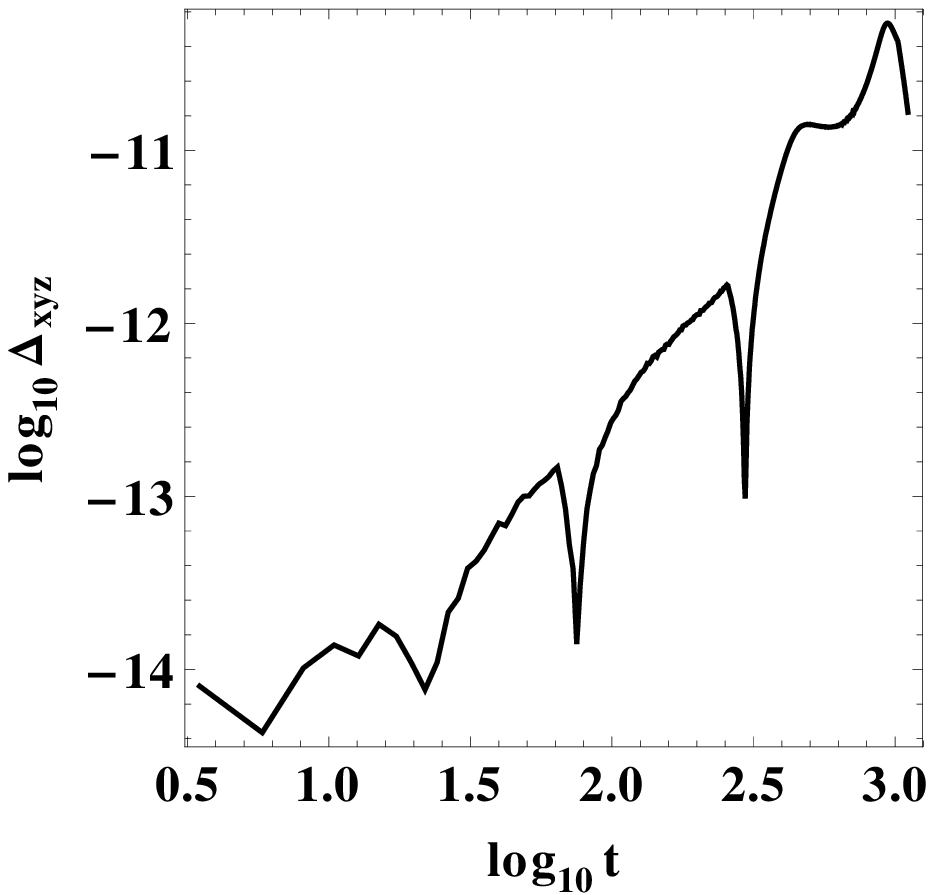}
 \includegraphics[width=0.25 \textwidth]{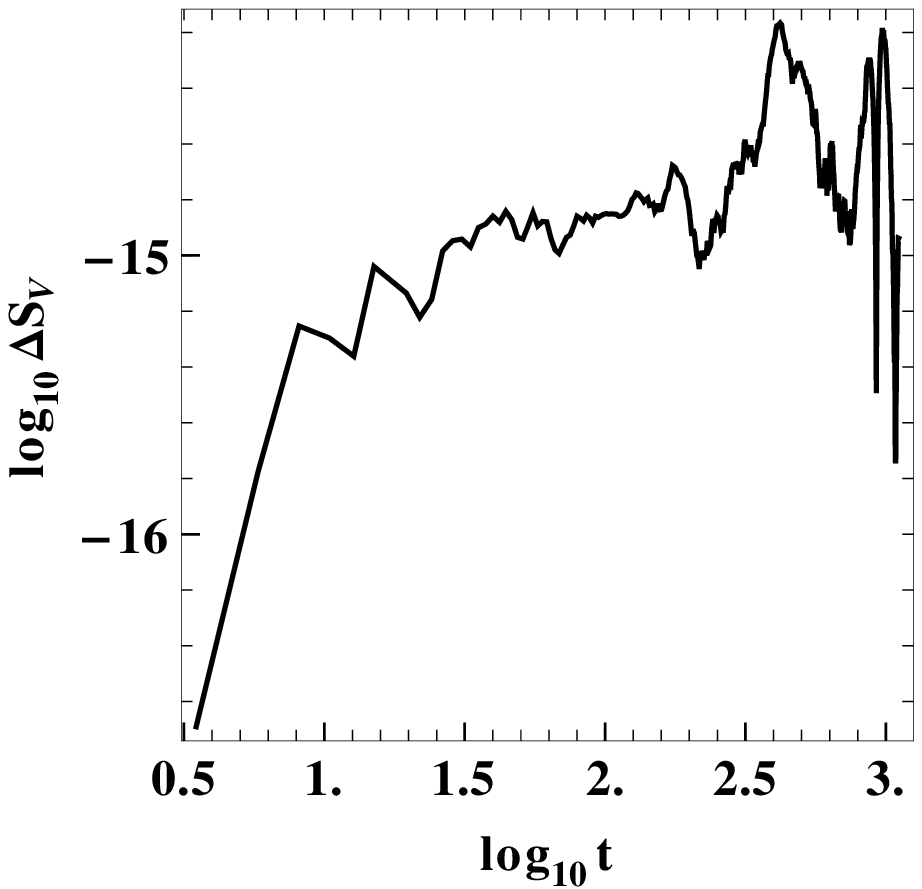}
 }
 \caption{The left panel shows the logarithm of the Euclidean distance in the
 configuration space between an orbit calculated with the MP equations and an orbit
 calculated with the Hamilton equations as a function of the coordinate time. For 
 the orbits we have used the initial conditions given in
 Fig.~\ref{Fig:ConSpCarta05m8}. The right panel shows the logarithm of the 
 Euclidean norm of the difference between the spin vectors of these two orbits
 as a function of the coordinate time.
 }
 \label{Fig:ConSpCartHa05m8}
 \end{figure}  

 By decreasing the measure of the particle's spin to the level of $S=10^{-8}$,  we
 get the initial setup given in Fig.~\ref{Fig:ConSpCarta05m8}. The Euclidean
 distance between the evolutions of the MP equations and the Hamilton equations
 (left panel of Fig.~\ref{Fig:ConSpCartHa05m8}) drops to a level which is near the
 precession of our simulations. Therefore, practically, the two orbits should not 
 discern. This seems to be the picture we get from the Euclidean norm of the
 difference between the spin vectors as well (right panel of
 Fig.~\ref{Fig:ConSpCartHa05m8}). Moreover, it is also evident that the distance
 between the two orbits does not exceed the diameter of the disc of centers
 of mass defined by the M\"{o}ller radius for the coordinate time we have computed. 
 Therefore, it is reasonable to say that the two orbits obtained by two different 
 formalisms do correspond to the same particle and thus infer that the Hamiltonian
 is indeed valid for small spin values.
 However, this picture might be a little bit illusive. The order of the spin is
 $S=10^{-8}$, and, thus, what we see in fact is that the relative difference, i.e.,
 $\Delta S_\textrm{v}/S \approx 10^{-8}$ is of the order of the spins' magnitude.
 In other words, in the spin space the evolution of the two orbits does not agree
 completely. The reason that in the configuration space the orbits appear to be
 identical, while in the spin space the agreement is not at the same level, is that
 we are in the geodesic limit, and the evolution of the orbits is almost 
 independent from the spins.  
 \begin{figure} [htp]
  \centerline{
  \includegraphics[width=0.25 \textwidth] {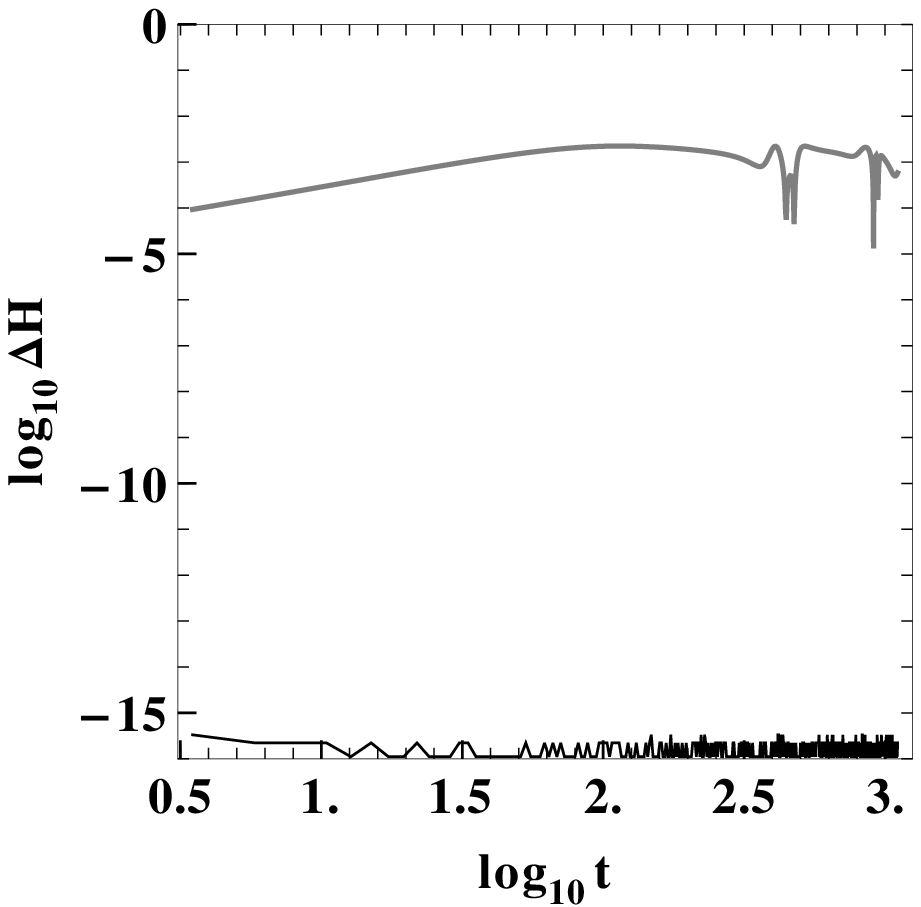}
 \includegraphics[width=0.25 \textwidth] {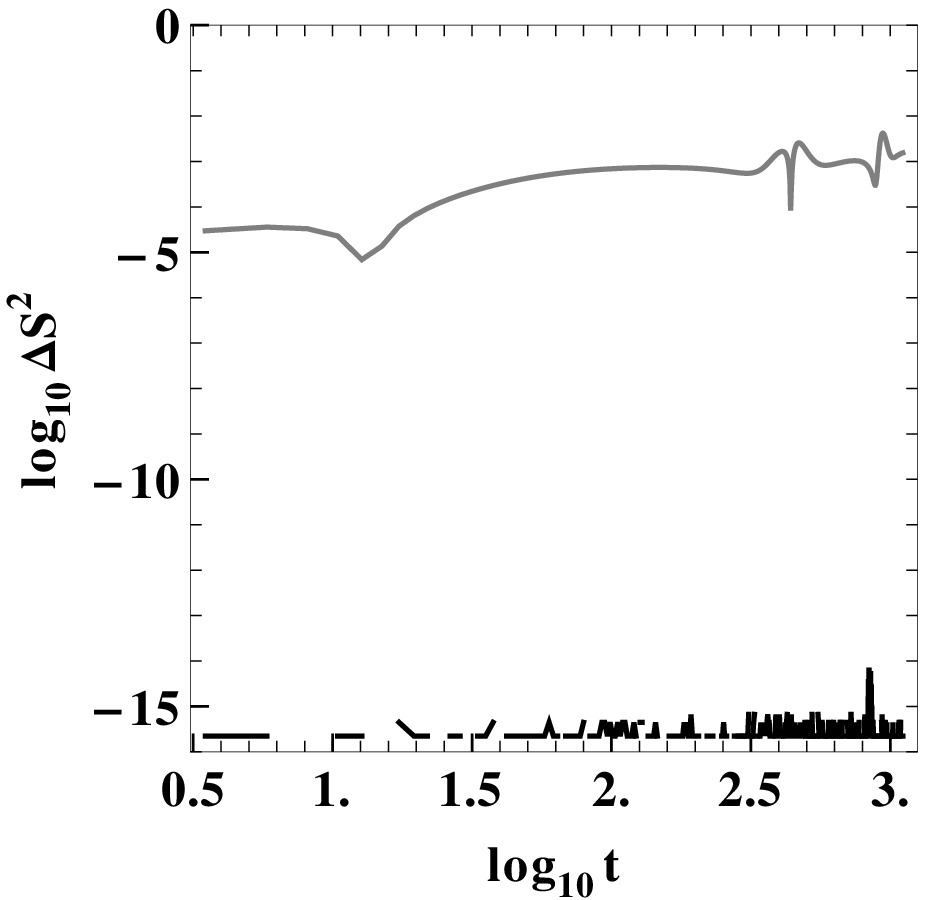}
 }
  \centerline{
  \includegraphics[width=0.25 \textwidth] {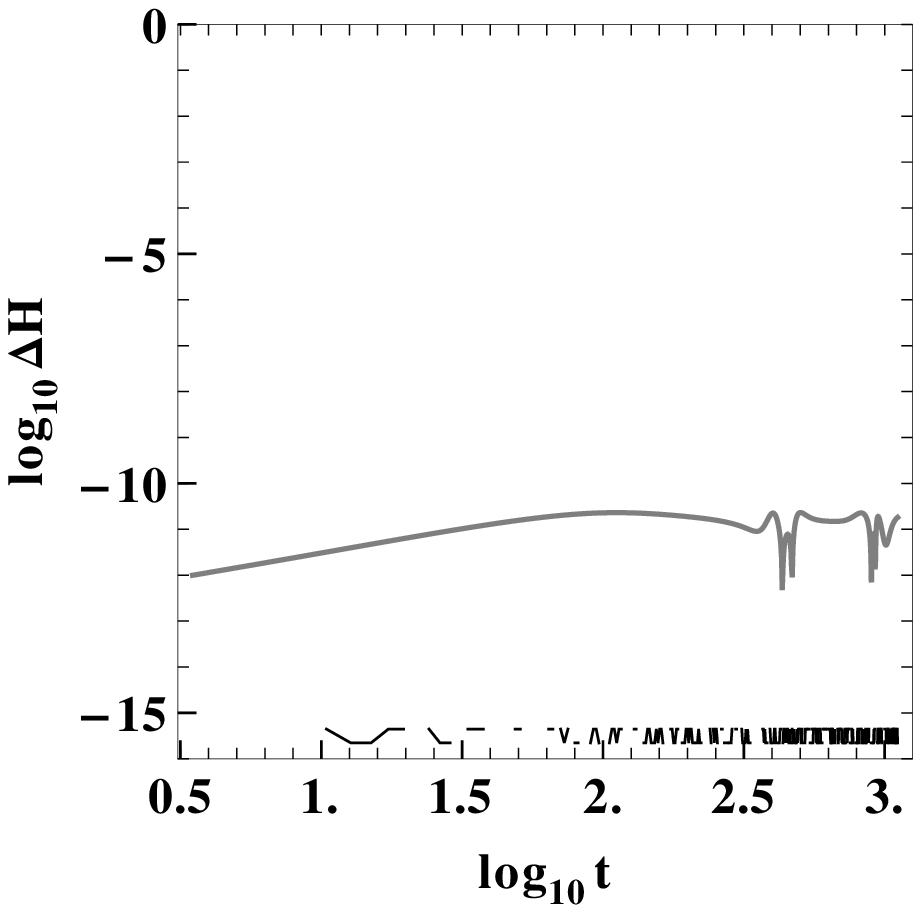}
 \includegraphics[width=0.25 \textwidth]{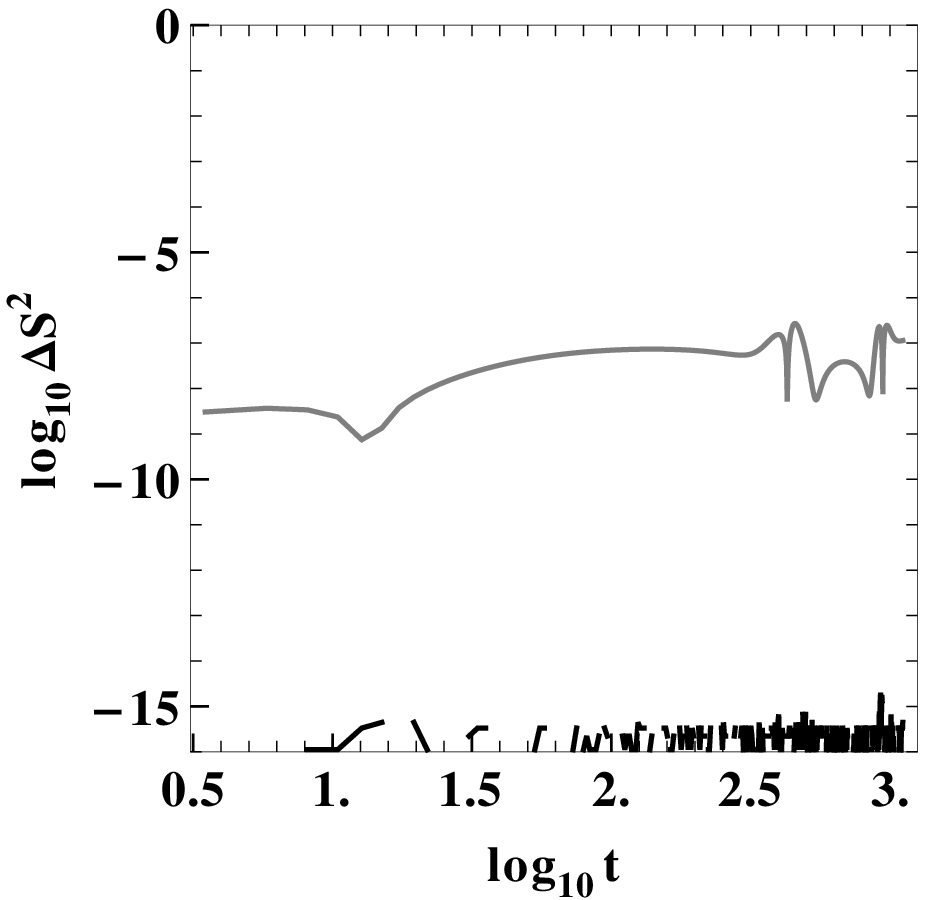}
 }
  \centerline{
  \includegraphics[width=0.25 \textwidth] {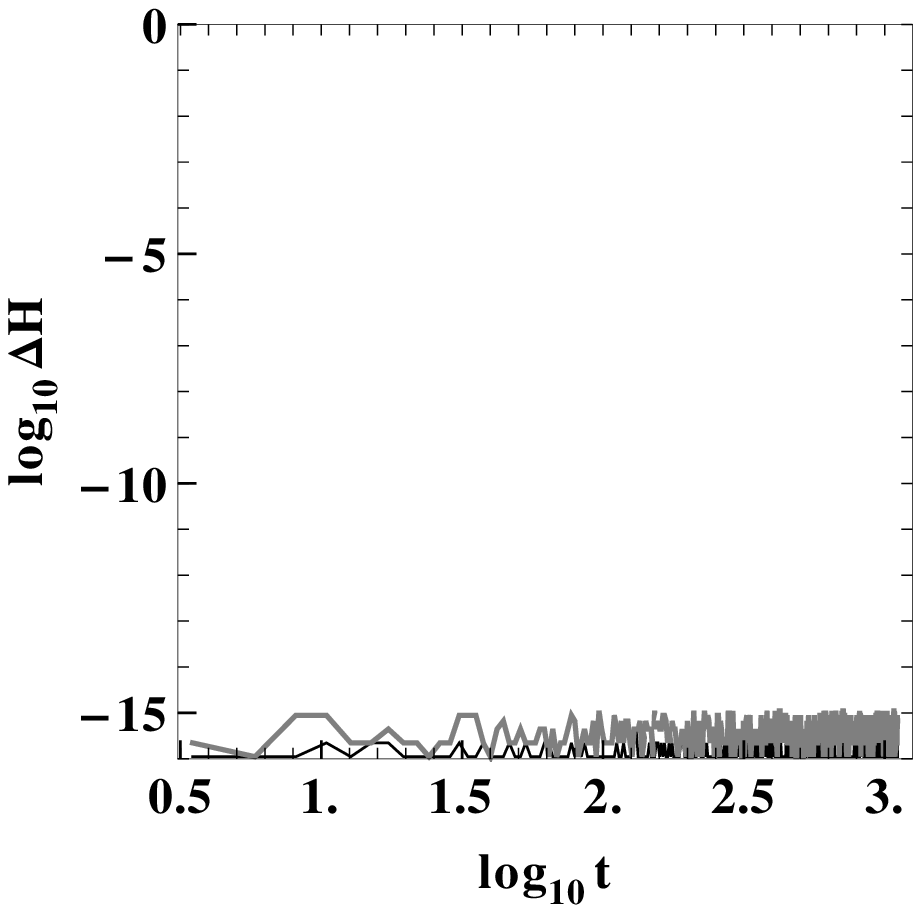}
 \includegraphics[width=0.25 \textwidth]{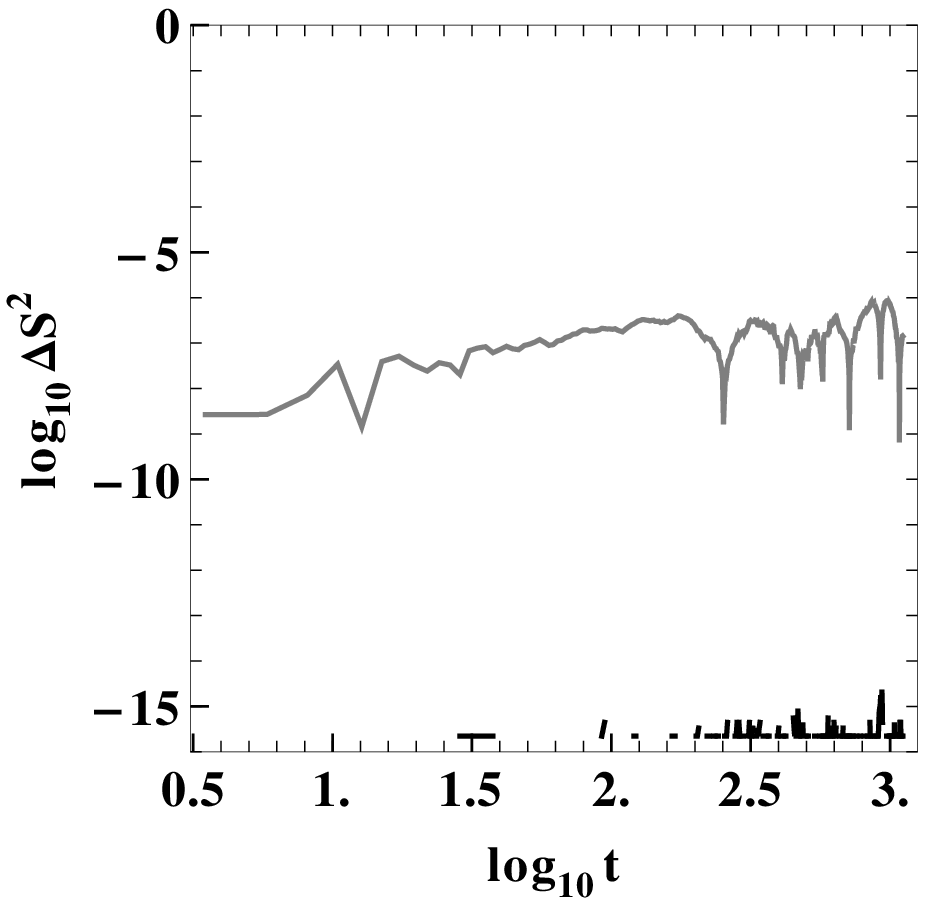}
 }
 \caption{ The top row of panels corresponds to the orbits of
 Fig.~\ref{Fig:ConSpCartHa05m0}, while the bottom row of panels corresponds to the
 orbits of Fig.~\ref{Fig:ConSpCartHa05m8}. The middle row of panels corresponds to
 initial conditions similar to Fig.~\ref{Fig:ConSpCarta05m0} only instead of
 spin measure $S=1$ we set $S=10^{-4}$. The gray lines represent the evolution
 of the MP equations, while the black lines represent the evolution of the Hamilton equations.
 The left column of panels shows the relative error in the preservation of the 
 Hamiltonian function, while the right shows the preservation of the spin.  
 }
 \label{Fig:HMPDS2Ha05m0m8}
 \end{figure} 

 The bottom row of Fig.~\ref{Fig:HMPDS2Ha05m0m8} supports the claim that when
 $S=10^{-8}$, we are at the geodesic limit, and the evolution does not depend on
 the spins. In the left panel of the bottom row in Fig.~\ref{Fig:HMPDS2Ha05m0m8},
 the relative errors of the Hamiltonian function,
 \begin{equation}
  \Delta H=\left|1-\frac{H(t)}{H(0)}\right|~~,\label{eq:RelErH}  
 \end{equation}
 lie at the computation precession level for both the MP orbit (gray line) 
 and the Hamiltonian orbit (black line), while the level of the relative
 error~\eqref{eq:RelErS2} in the measure of the spin vectors,
 \begin{equation}
     S^2= S_I S^I~~,\label{eq:SpinVec3M}
 \end{equation}
 is not as well preserved for the MP case (gray line) as for the Hamiltonian case
 (black line in the right panel of the bottom row in Fig.~\ref{Fig:HMPDS2Ha05m0m8}).
 Notice that, as stated above, in the case of the MP equation, we can get the value
 of the Hamiltonian function $H$ and of the square of the spin measure $S^2$ by
 transforming the set of variables $\{x^\mu,p^\mu,S^{\mu\nu}\}$ into the set
 $\{x^i,P^i,S^I\}$ and substituting the transformed set into 
 Eq.~\eqref{eq:HamSP} and Eq.~\eqref{eq:SpinVec3M} respectively.
 
 \subsection{Behavior of the constants of motion and scaling with the spin}

 When we raise the measure of the particle spin to $S=10^{-4}$, then the
 relative error of the MP spin (Eq.~\eqref{eq:SpinVec3M}) remains practically at
 the same level (gray line in the right panel of the middle row in
 Fig.~\ref{Fig:HMPDS2Ha05m0m8}) as in the $S=10^{-8}$ case. This does not hold for
 the relative error of the Hamiltonian function (gray line in the left panel of the
 middle row in Fig.~\ref{Fig:HMPDS2Ha05m0m8}) which is not at the computation
 precession level anymore. This shows that the motion is no longer in the geodesic 
 limit. However, both $\Delta S^2$ and $\Delta H$ for the MP orbit lie at
 acceptable levels, which shows that for this magnitude of the particle spin, the 
 MP equations and the Hamilton equations seem to be in agreement.

 This agreement breaks when $S=1$. The top row of Fig.~\ref{Fig:HMPDS2Ha05m0m8}
 shows that when $S=1$, the relative errors, $\Delta H$ and $\Delta S^2$ are
 at the same quite high level for the MP orbit. These relatively large values confirm the
 departure between the MP equations and the corresponding Hamiltonian that we see
 in Fig.~\ref{Fig:ConSpCartHa05m0}. 

 The black lines for all panels of Fig.~\ref{Fig:HMPDS2Ha05m0m8} are at the
 highest accuracy the computation accuracy allows, which means that apart
 from round-off error, the Gauss scheme we applied integrates accurately the system 
 of the Hamilton equations, but also that the interpolation scheme we applied to
 match the coordinate times works quite well.
 \begin{figure} [htp]
  \centerline{
  \includegraphics[width=0.25 \textwidth] {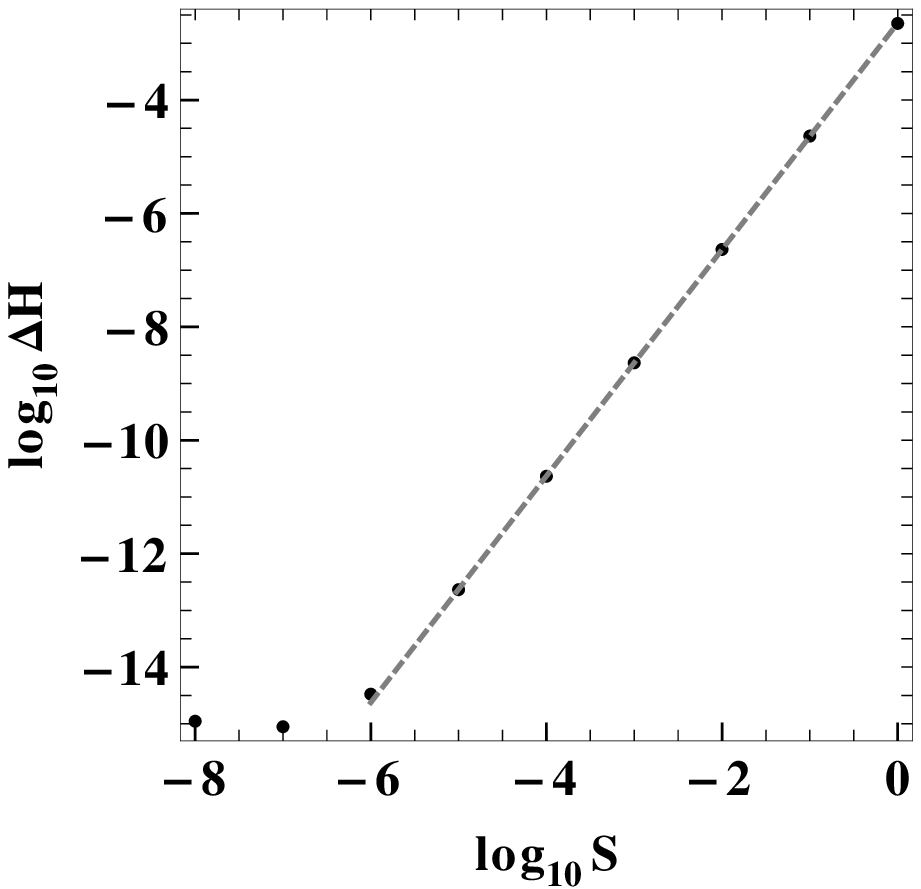}
 \includegraphics[width=0.25 \textwidth] {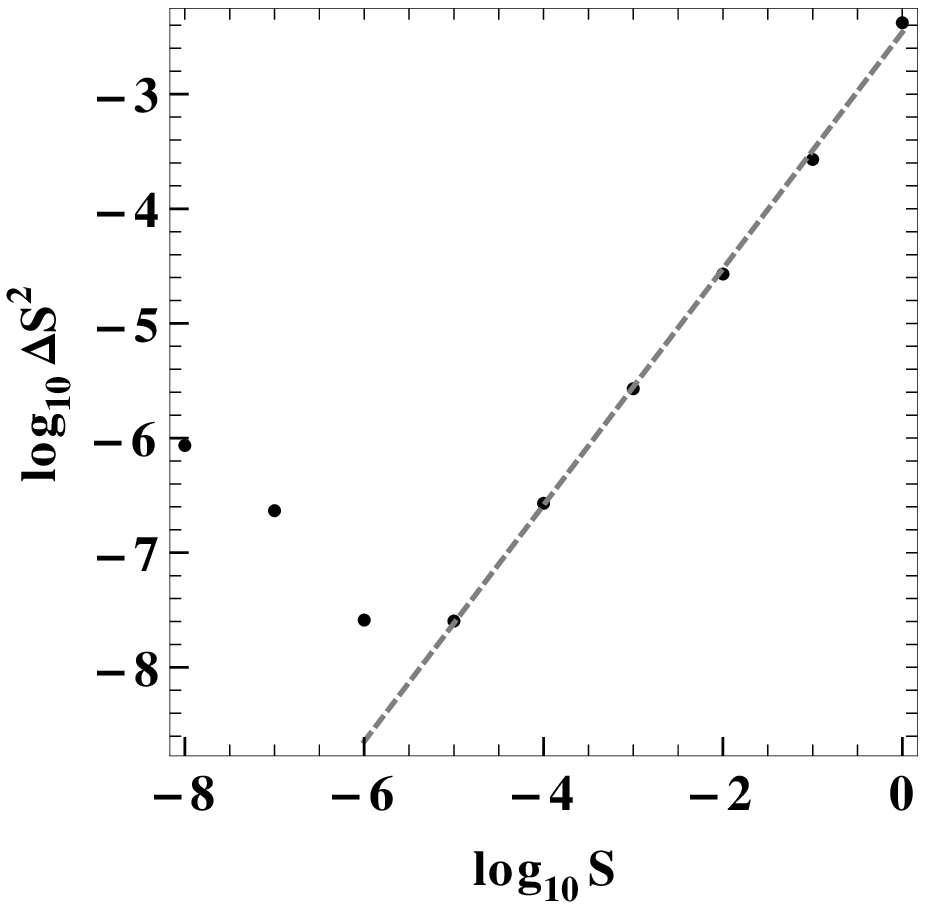}
 }
 \caption{The left panel shows the relative error of the Hamiltonian $\Delta H$
 of orbits evolved through the MP equations for different spin measures $S$ of the 
 particle, while the right panel shows the corresponding preservations of the 
 measure of the 3-vector $\Delta S^2$. The black dots correspond to the maximum 
 values of $\Delta H$, $\Delta S^2$, respectively, for each $S$. The dashed lines are 
 linear fits of the form  $\log_{10}{\Delta H}=a\log_{10}{S}+b$, and  
 \newline
 $\log_{10}{\Delta S^2}=c\log_{10}{S}+d$, respectively, for data with $S>10^{-6}$,
 where $a=1.9968\pm0.0015,~b=-2.644\pm0.004$, and \newline 
 $c=1.031\pm0.015,~d=-2.46\pm0.06$. 
 }
 \label{Fig:DH1DS2ScHa05}
 \end{figure} 

 As at the end of the previous Section, we can investigate the scaling of the
 constants of motion with the spin in more detail by taking the maxima of their 
 relative errors the MP equations, for different values of the measure of the
 particle's spin. The result is shown in Fig.~\ref{Fig:DH1DS2ScHa05}. Again, as
 in Fig.~\ref{Fig:MaSpSca05}, the precession of our computations and the
 scaling due to the spin measure shape the figure. We see a plateau at the left panel of
 Fig.~\ref{Fig:DH1DS2ScHa05} for $\Delta H$ due to the computational precession,
 while in the right panel of Fig.~\ref{Fig:DH1DS2ScHa05} we see that $\Delta S^2$
 increases, which is due to to the smallness of the spin components. Even if we had 
 applied a special integration scheme respecting these small quantities, this scheme
 could not follow below a threshold either. This threshold is in our case 
 $S=10^{-6}$. When the scaling with the spin dominates $(S>10^{-6})$, the linear
 fits show that $\Delta H \propto S^2$, while $\Delta S^2 \propto S$. These
 proportionalities are expected as we explain next.

 By construction the Hamiltonian function $H$ of a spinning particle is accurate up
 to linear order of the particle spin. Hence, when compared with the value of the 
 Hamiltonian function yielded from the evolution of the MP equations $H_{MP}(t)$,
 the difference between the two Hamiltonian function values should differ by terms
 of the order $O(S^2)$, i.e.,  
 \begin{equation}
   H_{MP}(t) \approx H(t)+O(S^2)~~.\label{eq:HamDif}
 \end{equation}
 However, since we have chosen the same initial conditions for both evolution 
 schemes, it holds that $H_{MP}(0)=H(0)$. Thus, the relative error~\eqref{eq:RelErH}
 for the MP equations reads
 \begin{eqnarray}
  \Delta H &=& \left|\frac{H_{MP}(t)-H_{MP}(0)}{H_{MP}(0)}\right| \nonumber \\
   &\approx& \left|\frac{H(t)-H(0)}{H(0)}+\frac{O(S^2)}{H(0)}\right|~~.
 \end{eqnarray}
 Since we do not expect the relative error $\frac{H(t)-H(0)}{H(0)}$ to depend on
 the value of the particle's spin, and this expectation is confirmed by the
 numerical findings (black lines in the left column of Fig.~\ref{Fig:HMPDS2Ha05m0m8}),
 we get the scaling $\Delta H \propto S^2$ of Fig.~\ref{Fig:DH1DS2ScHa05}.

 In order to explain the scaling of the relative error $\Delta S^2$, we use a
 similar way of reasoning. The preservation of the spin for the Hamiltonian
 formalism \eqref{eq:SpinVec3M} is $S^2$, thus a reasonable expectation is that for
 the MP case we should get values ${S^2}_{MP}(t)$ from  Eq.~\eqref{eq:SpinVec3M}
 which differ from the Hamiltonian case at order $O(S^3)$, i.e.,
 \begin{equation}
   {S^2}_{MP}(t) \approx S^2(t)+O(S^3)~~.\label{eq:SpinDif}
 \end{equation}
 Furthermore, we have ${S^2}_{MP}(0) = S^2(0)$. Thus, the relative
 error~\eqref{eq:RelErS2} for the MP equations reads
 \begin{eqnarray}
  \Delta S^2 &=& \left|\frac{{S^2}_{MP}(t)-{S^2}_{MP}(0)}{{S^2}_{MP}(0)}\right|
   \nonumber \\
   &\approx& \left|\frac{S^2(t)-S^2(0)}{S^2(0)}+\frac{O(S^3)}{S^2(0)}\right|~~,
 \end{eqnarray}
 which explains why we see that $\Delta S^2 \propto S$ in the right panel
 of Fig.~\ref{Fig:DH1DS2ScHa05}.

 If we take as a criterion the convergence of the  constants of motion shown
 in Fig.~\ref{Fig:DH1DS2ScHa05}, and consent that a relative error of the level of
 $10^{-6}$ is adequate to state that the different formalisms have converged, then
 from our comparison the Hamiltonian formalism is in agreement with the MP equations
 for the NW~SSC when the measure of the particle's spin is $S<10^{-4}$. When we
 reach $S \approx 10^{-6}$, the effect of the spin appears not to be important
 anymore, and the orbit evolves like a geodesic, i.e., it does not depend on the
 spin. 

\section{Conclusions} \label{sec:con}

 We have compared the evolutions of a spinning test particle in Kerr spacetime
 governed by different equations of motion. We first evolved the orbits prescribed 
 by the MP equations, once supplemented by the Tulczyjew SSC and once by the
 Newton-Wigner SSC. Our simulations indicate a linear in the spin scaling of
 the difference between the respective orbits. We also found that, in the case of
 the NW~SSC, the four-momentum is conserved up to linear order in the square of the
 test particle's spin, i.e $\Delta \mu^2 \propto S^2$. In a second series of
 experiments we compared orbits given by the MP equations plus NW~SSC with orbits
 obtained via the Hamiltonian formalism of~\cite{Barausse09}. Here, too, the
 difference between the respective orbits, which is quite significant for large
 spins of the order of one, decreases linearly as a function of the square of the 
 test particle's spin, i.e. $\Delta H \propto S^2$, which agrees with the analysis
 given in \cite{Barausse09}. According to our analysis, the Hamiltonian formalism
 of the spinning particle appears to be relevant in the range $10^{-6}<S<10^{-4}$.
 For values of the spin smaller than $10^{-6}$ we can ignore the part of the
 Hamiltonian describing the spin evolution and keep the non-spinning part, and
 for spin values greater than $10^{-4}$, our numerical results show that the
 Hamiltonian formalism is not in good agreement with the MP equations. Anyhow,
 the aforementioned range, where the Hamiltonian formalism is relevant, is
 appropriate for astrophysical binary systems of extreme mass ratio. Moreover, as
 our simulations showed that the CPU effort for the Hamilton equations of motion
 is far smaller than the computational cost for the MP equations, we find
 appropriate the use of these equations for simulations of test particles with
 small spins. When, in addition, favorable numerical methods, such as the one
 presented in this work, are applied, reliable results can be obtained within a
 short period of time.

\begin{acknowledgments}
 This work was supported by the DFG grant SFB/Transregio 7, by the DFG Research 
 Training Group 1620 ``Models of Gravity '' and by the ``Centre for Quantum
 Engineering and Space-Time Research (QUEST)''. We specially would like to thank
 Oldrich Semer\'ak for his useful suggestions and remarks.

\end{acknowledgments}    

 \appendix

 \section{Numerical integration of the MP equations} \label{sec:NumIntMP}

 Seen from a numerical point of view, the initial value
 problem~\eqref{eqn-initial-value-problem-MP} reads
 \begin{align}
  \frac{\mathrm d\mathbf y}{\mathrm d\tau}&=f(\mathbf y)~~,\\
  \mathbf y(\tau=0)&=\mathbf y_0~~.
 \end{align}
 with $\mathbf y=(t,r,...,S^{\theta\phi},S^{\theta\theta})^T\in\mathbb R^{24}$ and 
 $f:\mathbb R^{24}\to\mathbb R^{24}$. If this system was of Hamiltonian canonical
 form, symplectic integration schemes would be the most natural choice for their
 numerical solution. They almost exactly preserve a differential equation's
 constants of motion and, unless for standard integration schemes, their overall
 numerical error grows only slowly as a function of the total integration time even
 for larger step sizes. Therefore, simulations over long time spans can be carried
 out efficiently. Unfortunately, the MP equations are not of Hamiltonian canonical
 form. But, they can be interpreted as the Euler-Lagrange equations of a suitable 
 Lagrangian action, see, e.g.,~\cite{Westpfahl69,Bailey75,Porto06}. What then saves
 the day is that the flow of symplectic integration schemes can be interpreted as 
 the solution of the Euler-Lagrange equations of a discretization of the Lagrangian
 action. Schemes with this property are called \textit{variational integrators}
 and they only rely on the existence of a Lagrangian structure for their favorable
 behavior. For example they are known to exactly preserve an equation of motion's
 first integrals which are quadratic in the phase space variables. This implies
 that a variational integration scheme applied to the MP equations with T~SSC will
 conserve the four-momentum $\mu^2$ and the spin length $S^2$ up to numerical
 round-off errors. An extensive discussion  of this topic can be found in the
 monograph~\cite{hairerlubichwanner}, chapter VI.6. One prominent example of 
 variational integrators is Gauss Runge-Kutta methods which have been shown to be
 the most efficient and accurate integrators in many general relativistic
 applications, see, e.g.,~\cite{Seyrich12,Seyrich13}. Motivated by these
 results, we choose this kind of variational integrator for the solution of the MP
 equations. Here we briefly summarize some of their properties.
 
 An $s$-stage Gauss Runge-Kutta scheme is a collocation method, i.e. an implicit
 Runge-Kutta scheme
 \begin{align}
  \mathbf y_{n+1} &= \mathbf y_n+h\sum_{i=1}^sb_if(\mathbf Y_i)\label{def-rk-y_n+1}~~,\\
  \mathbf Y_i &=\mathbf y_n+ h\sum_{j=1}^sa_{ij}f(\mathbf Y_j),\quad i=1,...,s\label{def-rk-Y_i}~~,
 \end{align}
 with coefficients
 \begin{align}
    a_{ij}=\int_0^{c_i}l_j(t)dt~~,\\
    b_j=\int_0^1l_i(t)dt~~,
  \end{align}
  where the stages $c_1,...,c_s$ are chosen as 
 \begin{align}
    c_i=\frac 1 2(1+\tilde c_i)~~,
 \end{align}
  with $\tilde c_i$ being the roots of the Legendre-polynomial of degree $s$. Here,
  $h$ denotes the time step size, $Y_i$, $i=1,...,s$, are the so-called inner
  stage values and $\mathbf y_n$ denotes the numerical approximation to the
  solution $\mathbf y$ at time $\tau=nh$. The functions $l_i(t)$ are the
  Lagrange-polynomials of degree $s$,
 \begin{align}
  l_i(t)=\prod_{i\neq j}\frac{t-c_j}{c_i-c_j}~~.
 \end{align}
 Gauss Runge-Kutta methods have a convergence order $\mathcal O(h^{2s})$ which is
 the highest possible order among collocation schemes,
 e.g.,~\cite{hairernorsettwanner}. When integrating a time step with a Gauss
 Runge-Kutta scheme, one first solves the system of implicit
 equations~\eqref{def-rk-Y_i} via a fixed-point iteration
 \begin{align}
  \mathbf Y^{k+1}_i &=\mathbf y_n+ h\sum_{j=1}^sa_{ij}f(\mathbf Y^k_j)~~.
 \end{align}
 This, of course, requires more calculations per time step than an explicit scheme 
 with the same number of stages. But, this extra effort is more than offset by the
 high accuracy of Gauss collocation methods which allows us to apply them with a much
 larger step size. Detailed information on their implementation is given
 in~\cite{Seyrich13}, Sec.~7, and \cite{hairerlubichwanner},
 chapters VIII.5 and VIII.6.
 
\begin{figure} [htp]
  \centering
  \includegraphics[width=0.5\textwidth]{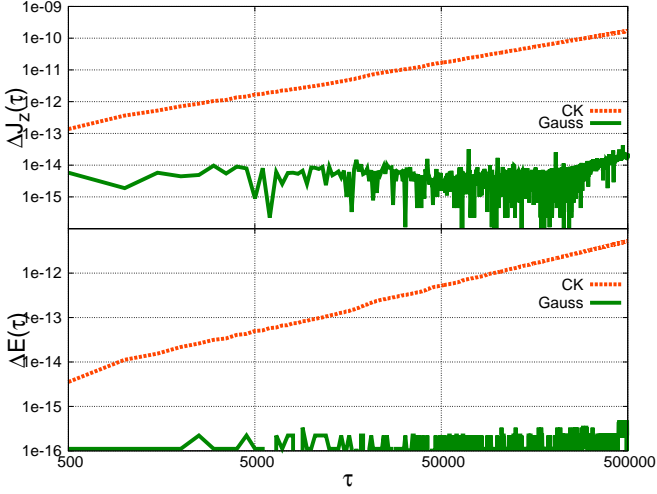}
  \caption{The relative error of the $z$ angular momentum, $\Delta J_z$,
  (top panel) and the relative  error of the energy, $\Delta E$, (bottom panel)
  against integration time $\tau$ for the $4$-stage Gauss scheme with step size 
  $h=1$ and the $5$-th order Cash-Karp scheme with step size $h=0.1$ applied to the
  initial value problem~\eqref{eqn-initial-value-problem-MP} with initial data as
  stated in the text. CPU-time was $214.1\mathrm s$ for the Gauss Runge-Kutta
  scheme and $422.7\mathrm s$ for the Cash-Karp scheme.}
  \label{fig-comp-Gauss-CK-MP}
 \end{figure}

 To illustrate the favorable behavior of Gauss collocation methods, we compare the
 performance of a $4$-stage scheme with step size $h=1$ and a standard $5$-th order
 explicit Cash-Karp scheme as proposed in~\cite{nr} with a step size $h=0.1$, when
 applied to the MP equations with T~SSC and initial data given by $E=0.95$, $J_z=3.0$,
 $S=1$, $M=1$ $\mu=1$, $a=0.9$, $r=6.7$, $\theta=\frac\pi2+0.1$, $p_r=0.1$, $S_r=0.1$,
 $S_\theta=0.01$. 
 In Fig.~\ref{fig-comp-Gauss-CK-MP}, we plot for both integrators 
 the relative error in the energy,
 \begin{align}
 \Delta E(\tau)=\frac{|E(\tau)-E(0)|}{|E(0)|}~~,
 \end{align}
 and the corresponding relative error in the $z$ angular momentum as a function
 of integration time $\tau$. We observe that the Gauss Runge-Kutta method, which 
 is also faster, gives much more precise results.

 An additional obstacle for simulations in the NW~SSC case is that the tangential
 velocity $v^\mu$ is only given implicitly by Eq.~\eqref{eq:NWTanV}. (N.b.: Apart 
 from the apparent $v^\nu$ in the first term on the right hand side, the  covariant
 derivative of $\omega_\nu$ implies a linear dependence on $v^\nu$ in the second 
 term on the rhs as well, i.e., $\displaystyle \frac{D~\omega_\nu}{d \tau}=
 \dot{\omega}_\nu-\Gamma^\kappa_{\nu\mu}\omega_\kappa v^\mu$.)  Setting
 $\vec v:=(v^t,v^r,v^\phi,v^\theta)^T\in\mathbb R^4$, the implicit equation for
 $v^\mu$ is qualitatively given by
 \begin{align}\label{eqn-v-implicit}
  \vec v=A(x^\mu,p^\mu,S^{\mu\nu})\vec v
 \end{align}
 for a certain matrix $A\in\mathbb R^{4\times4}$. Theoretically there are two
 possibilities to cope with the implicitness in the velocities which we will
 describe now.
 \begin{itemize}
 \item Denoting the first four components of $Y_i$ and $f(Y_i)$ by $Y_i^x$ and
  $f^x(Y_i)$, and the other components by $Y_i^p$, $Y_i^S$, $f^p(Y_i)$, and
  $f^S(Y_i)$ we can augment the system of implicit equations~\eqref{def-rk-Y_i}
  by adding the implicitly given quantity $\vec v_i$ which denotes the tangential 
  velocity $v^\mu$ at the inner stage $Y_i$. This yields the system
 \begin{align}
  \begin{pmatrix}\vec v_i\\\mathbf Y^x_i\\\mathbf Y^p_i\\\mathbf Y^S_i\end{pmatrix} &=
  \begin{pmatrix}A(Y_i^x,Y_i^p,Y_i^S)\vec v_i\\\mathbf y^x_n+ h\sum_{j=1}^sa_{ij}\vec v_i\\
   \mathbf y^p_n+ h\sum_{j=1}^sa_{ij}f^p(Y_i^x,Y_i^p,Y_i^S,\vec v_i)\\
    \mathbf y^S_n+ h\sum_{j=1}^sa_{ij}f^S(Y_i^x,Y_i^p,Y_i^S,\vec v_i)\end{pmatrix}~~,\nonumber\\
   \quad i=1,...,s~~,
 \end{align}
 to which, again, a fixed-point iteration can be applied. However, for this iteration
 to converge, it needs to satisfy
   \begin{align}
    \vert|\begin{pmatrix}\vec v^{k+2}_i\\Y^{k+2}_i\end{pmatrix}-\begin{pmatrix}\vec v^{k+1}_i\\Y^{k+1}_i\end{pmatrix}\vert|\le
    \vert|\begin{pmatrix}\vec v^{k+1}_i\\Y^{k+1}_i\end{pmatrix}-\begin{pmatrix}\vec v^k_i\\Y^k_i\end{pmatrix}\vert|~~,
   \end{align}
 which cannot be guaranteed when $A(Y_i^x,Y_i^p,Y_i^S)$ is of large norm.
 Numerical tests have shown that there are indeed problems with the convergence.
 Hence, for all its conceptual beauty, the approach of an augmented implicit system
 is of no practical use.

 \item With $I$ denoting the $4\times4$ identity matrix, we can rewrite the
 implicit equation for the velocities~\eqref{eqn-v-implicit} as
 \begin{align}
   0=(I-A)\vec v=:B\vec v~~.
 \end{align}
 Thus, from an algebraical point of view, the vector consisting of the components
 of the $4$-velocity is an element of the nullspace $\operatorname{Ker}(B)$ of the
 matrix B which here is a one-dimensional subspace. Consequently, we can determine 
 the tangential velocity at an internal stage by the following procedure
 \begin{enumerate}
  \item[1.] Calculate \newline $B(Y_i^x,Y_i^p,Y_i^S)=I-A(Y_i^x,Y_i^p,Y_i^S)$.
  \item[2.] Calculate the singular-value-decomposition of $B$, i.e.,
  \begin{align}
    B=U\Sigma V^T~~,
  \end{align}
  with $\Sigma=\operatorname{diag}(\sigma_1,\sigma_2,\sigma_3,\sigma_4)$ and
  $U^TU=V^TV=\delta_{ij}$, $i,j=1,...4$ (For more information on the singular 
  value decomposition, see, e.g.,~\cite{nr}, chapter 2.6). The nullspace
  of B is then spanned by the column of the orthonormal matrix $V_{.,i}$ that
  corresponds to the only singular value $\sigma_i$ which is equal to $0$.
  \item[3.] The tangential velocity is now obtained by renormalizing $V_{.,i}$ in
  order to have $v^\mu v_\mu=-1$. 
 \end{enumerate}
 This procedure is very robust and the computational cost for the calculation of
 the matrix $B$ and the singular value decomposition is far less than the
 computational cost for the calculation of the other quantities which are needed
 anyway. This could be confirmed experimentally when comparing CPU times for
 simulations with T~SSC and NW~SSC for similar initial values. For all the
 simulations done in the preparation for this work, the CPU times in the NW~SSC
 case were only slightly higher than those for the T~SSC case where the
 velocities could be determined explicitly via Eq.~\eqref{eq:TTanVA}.
\end{itemize}
 Last, we turn to the numerical integration of the Hamiltonian formalism in the next
 section.

 \section{Numerical integration of the Hamiltonian equations} \label{sec:NumIntHam}

\begin{figure} [htp]
  \centering
  \includegraphics[width=0.5\textwidth]{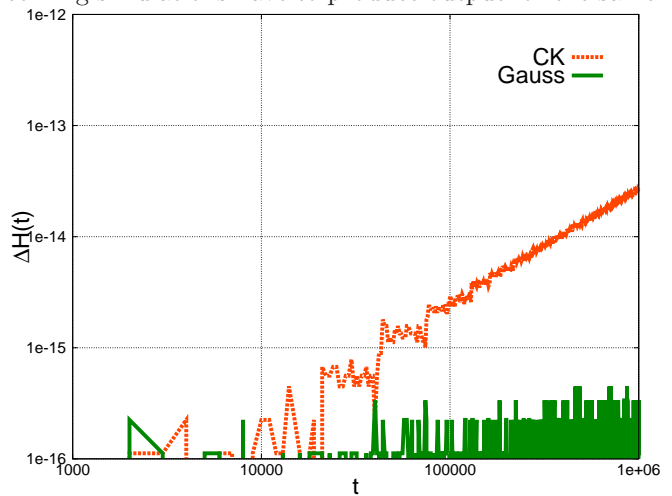}
  \caption{The relative error of the Hamiltonian, $\Delta H$ against integration
  time $t$ for the $4$-stage Gauss scheme with step size $h=2$ and the $5$-th order
  Cash-Karp scheme with step size $h=0.2$ applied to the initial value
  problem~\eqref{eqn-def-poisson-sys} with initial data as stated in the text.
  CPU-time was $7.83\mathrm s$ for the Gauss Runge-Kutta scheme and $24,7\mathrm s$
  for the Cash-Karp scheme.}
  \label{fig-comp-Gauss-CK-Ham}
 \end{figure}

 The Hamiltonian equations considered in this study have a so-called
 \textit{Poisson structure}, that is, with
 $\mathbf y=(P_r,P_\theta,P_\phi,r,\theta,\phi,S_1,S_2,S_3)^T\in\mathbb R^9$, they 
 can be written as
 \begin{align}\label{eqn-def-poisson-sys}
  \dot{\mathbf y}=B(\mathbf y)\grad H(\mathbf y)~~,
 \end{align}
 where $B:\mathbb R^9\to\mathbb R^{9\times9}$ is a skew-symmetric matrix-valued
 function. In our case, this function $B(\mathbf y)$ is given by
 \begin{align}
  B(\mathbf y)&=\begin{pmatrix}0&-I_{3\times3}&0\\I_{3\times3}&0&0\\0&0&B_1(\mathbf y)\end{pmatrix}~~,
  \end{align}
  with
 \begin{align}
  I_{3\times3}&=\begin{pmatrix}1&0&0\\0&1&0\\0&0&1\end{pmatrix}~~,\\
  B_1(\mathbf y)&=\begin{pmatrix}0&-S_3&S_2\\S_3&0&-S_1\\-S_2&S_1&0\end{pmatrix}~~.
 \end{align}
 For such $B(\mathbf y)$, there exists a smooth transformation to new coordinates
 $\mathbf z$, for which the equations of motion are of symplectic form
 \begin{align}\label{eqn-def-sympl-sys}
  \dot{\mathbf z}&=J^{-1}\grad H(\mathbf z)~~,\\
  J&=\begin{pmatrix}0&I_{4\times4}\\-I_{4\times4}&0\end{pmatrix}~~,
\end{align}
 see~\cite{wuxie,Seyrich13}. The idea how to find this transformation is based on
 the conservation of the spin length $S=\sqrt{S_1^2+S_2^2+S_3^2}$ by the 
 eqs.~\eqref{eqn-def-poisson-sys}. Thus, the three dimensional spin
 $\mathbf S=(S_1,S_2,S_3)^T$ can be given as a function of two variables
 $\alpha$ and $\xi$ via
 \begin{align}
  \mathbf S=S\begin{pmatrix}\sqrt{1-\xi^2}\cos(\alpha)\\\sqrt{1-\xi^2}\sin(\alpha)\\\xi\end{pmatrix}~~.
 \end{align}
 One can then show that 
 \begin{align}
  \dot\xi&=-\pd H\alpha~~,\\
  \dot\alpha&=\pd H\xi~~
 \end{align}
 hold, see, e.g.~\cite{Seyrich13}. Hence, for the variables
 $\mathbf z=(P_r,P_\theta,P_\phi,\xi,r,\theta,\phi,\alpha)$, the equations of
 motion indeed take the form~\eqref{eqn-def-sympl-sys}. Whenever a system can be
 smoothly transformed to symplectic form, it can be evolved by symplectic
 integration schemes. Therefore, for our studies of the Hamiltonian formalism
 of~\cite{Barausse09}, we follow~\cite{Seyrich13} and use Gauss Runge-Kutta schemes
 which have already been presented in the last section~\footnote{As opposed to the
 approach in~\cite{Seyrich13} we did not bother to rewrite the system in the
 variables $z$, because in the present case the additional cost of the one extra 
 variable is negligible in comparison to the other computational effort.}. In order
 to show their favorable behavior, we evolve the Hamiltonian system for initial
 data $M=1$, $m=1$, $a=\frac1{10}$, $r=15$, $\theta=\frac\pi2$, $\phi=0$, $P_r=0$, 
 $P_\theta=3.69336$, $P_\phi=J_z=3.8$, $S_1=\frac1{\sqrt{2}}$,
 $S_2=\frac1{\sqrt{3}}$, $S_3=\frac1{\sqrt{6}}$ and plot, in Fig.~\ref{fig-comp-Gauss-CK-Ham},
 the relative error of the Hamiltonian~\eqref{eq:RelErH} once for the
 Gauss Runge-Kutta method with $s=4$ inner stages and once for the $5$th order
 explicit Cash-Karp scheme. For the explicit method we observe a linear growth in
 the error while there is no significant error during the whole simulation for the
 Gauss scheme. This is in spite of the latter's much smaller CPU time. With regard
 to the computational effort, we also notice that it is much smaller than in the
 case of the full MP equations, although both cases were tested on the same machine.
 This gives another practical reason to consider the Hamiltonian approximation. 
 
 In our comparison of the orbits given by the MP equations with those of the
 Hamiltonian formalism, the concerning simulations have to produce output for the 
 same coordinate times. To avoid having to reformulate the MP equations for the
 coordinate time as evolution parameter, we proceed as follows. In the simulation
 of the MP equations, output is produced at uniform distances in the evolution
 parameter proper time. The output also comprises the corresponding coordinate
 times. These are then fed as input to the Hamiltonian simulations -for example
 under the name $t_\text{output required}$.  Now, if in the simulation with uniform
 steps in the evolution parameter coordinate time $t$, between times $t_i$ and
 $t_{i+1}$ say, one passes one of the prescribed times for which output is required,
 $t_\text{output required}$,  one can take use of the interpolation property
 of the collocation schemes to comfortably obtain output at no computational extra
 cost. It is well known that the interpolation polynomial $\mathbf u(t)$ through 
 the points $(0,\mathbf y_n)$, $(c_i,\mathbf Y_i)$, $i=1,...,s$, stays
 $\mathcal O(h^s)$ close to the exact solution of the equation of motion, and,
 hence, also to the numerical calculated trajectory,
 see,~e.g.,~\cite{hairernorsettwanner}. We thus only have to evaluate
 $\mathbf u(t)$ at time $t_\text{output required}-t_i$ which yields an
 approximation of the solution at time $t_\text{output required}$ which is exact up
 to an error of $\mathcal O(h^s)$. The interpolation polynomial itself can be
 calculated very quickly with the so-called Horner scheme

\begin{widetext}
 \begin{align}
  &\mathbf u(t)=\mathbf y_i+(t-0)\left(\delta^1[0,hc_1]+(t-hc_1)\left(\delta^2[0,hc_1,hc_2]\right.\right.
  \left.\left.+(t-hc_2)\left(...(t-hc_{s-1})\delta^s[0,hc_1,...,hc_s]\right)...\right)\right)~~,\nonumber\\
  &\delta^1[0,hc_1]=\frac{\mathbf Y_1-\mathbf y_i}{hc_1-0}~~,\nonumber\\
  &\delta^k[0,hc_1,...,hc_k]= 
  \frac{\delta^{k-1}[hc_1,...,hc_k]-\delta^{k-1}[0,hc_1,...,hc_{k-1}]}{hc_k-0}~~.
 \end{align}
\end{widetext}

 \begin{figure} [htp]
  \centering
  \includegraphics[width=0.5\textwidth]{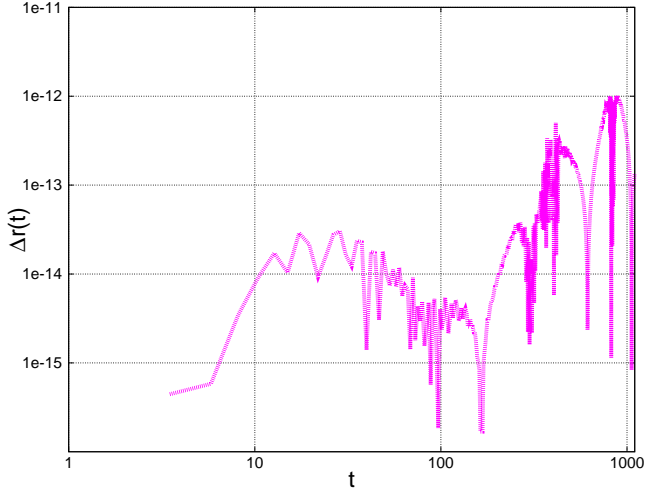}
  \caption{The relative difference, $\Delta r$, between the radial distance
  calculated with the interpolation method and the radial distance calculated via 
   the cumbersome method with extra integration steps plotted against output time 
  $t$.}
  \label{fig-comp-int-cumb}
 \end{figure}

 The more intricate way of producing output at the desired times would be the following:
 \begin{itemize}
 \item{} When having passed an output time $t_\text{output required}$ between $t_i$
 and $t_{i+1}$, go back to $t_i$.
 \item{} Change $h\to h_\text{new}=t_\text{output required}-t_i$.
 \item{} Evolve the system until $t=t_\text{output required}$ with step size
 $h_\text{new}$ and produce output.
 \item{} Go back to $t_i$ and go on integrating with step size $h$. (Note that this
 is necessary as the scheme would lose
 its symplectic structure when applied with different step sizes,
 see, e.g.~\cite{hairerlubichwanner}, chapter VIII.)
 \end{itemize}

 In order to illustrate that this cumbersome procedure is not worth the additional
 effort, we again consider the data which yielded Fig.~\ref{Fig:ConSpCartHa05m0}
 and, for every coordinate time $t$, for which $\Delta_{xyz}$ was plotted in the
 central panel of that figure, we plot the relative difference in the radial
 distance at those times between the interpolation method and the cumbersome method,
 \begin{align}
  \Delta r(t)=\frac{|r_\text{interpolation}(t)-r_\text{cumbersome}(t)|}{r}~~.
 \end{align}
 In Fig.~\ref{fig-comp-int-cumb}, we can observe that the difference is negligible.

\end{document}